\documentclass[graybox]{svmult}

\usepackage{mathptmx}       % selects Times Roman as basic font
\usepackage{helvet}         % selects Helvetica as sans-serif font
\usepackage{courier}        % selects Courier as typewriter font
\usepackage{makeidx}         % allows index generation
\usepackage{graphicx}        % standard LaTeX graphics tool
\usepackage{multicol}        % used for the two-column index
\usepackage[bottom]{footmisc}% places footnotes at page bottom

\makeindex             % used for the subject index
\usepackage{amsfonts}
\usepackage{amsmath}
\usepackage{subfigure}
\usepackage{keyval}
\usepackage{color}
\newcommand{\real}{{\mathbb R}}
\newcommand{\reald}{\real^{\rm{d}}}

\newcommand{\eps}{\varepsilon}
\newcommand{\vr}{\mathcal{V}}
\newcommand{\n}{{\mathbb N}}
\newcommand{\pl}{{\rm L}}

\usepackage[square,numbers]{natbib}

\makeindex

\begin{document}

\title*{Topological Data Analysis of {\it Clostridioides} {\it difficile} Infection and Fecal Microbiota Transplantation}
\titlerunning{Topological Data Analysis of CDI and FMT} 
\author{Pavel Petrov\ \ Stephen T. Rush\ \ Zhichun Zhai\ \ Christine H. Lee \ \ Peter T. Kim\ \ and \ \ Giseon Heo}
\authorrunning{P Petrov ST Rush Z Zhai CH Lee PT Kim G Heo}
\institute{Pavel Petrov, MSc \at Consultant at MNP, 21 West Hastings Street
Vancouver , BC.\ V6E 0C3 Canada.
\and Stephen Rush, PhD\at Department of Medical Sciences, University of \"{O}rebro, \"{O}rebro, 702 81, Sweden. \email{stephen.rush@oru.se}
\and  Zhichun Zhai,  PhD \at  Department of Mathematical and Statistical Sciences, University of Alberta, T6G 2G1 Canada.
\email{zhichun1@ualberta.ca}
\and  Christine Lee, MD FRCPC.\ \at Department of Microbiology, Royal Jubilee Hosptial, Victoria, BC, V8R 1J8, Canada. \email{christine.lee@viha.ca}
\and  Peter Kim, PhD\at Department of Mathematics \& Statistics, University of Guelph, Guelph, Ontario, N1G 2W1 Canada.  \email{pkim@uoguelph.ca}
\and Giseon Heo, PhD\at School of Dentistry; Department of Mathematical and Statistical Sciences, University of Alberta, Edmonton, T6G 1C9  Canada.
\email{gheo@ualberta.ca}
}

\maketitle

\abstract{
Computational topologists recently  developed a method, called persistent homology to
analyze data presented in terms of similarity or dissimilarity.  Indeed,
persistent homology studies the evolution of topological features in terms of a single index, and is
able to capture higher order features beyond the usual clustering techniques.    
There are three descriptive statistics of persistent homology, namely barcode, persistence diagram and more recently, persistence landscape.
Persistence landscape  is useful for statistical inference as it belongs to a space  of $p-$integrable  functions, a separable Banach space.
We apply tools in  both computational topology and statistics to  DNA sequences taken from {\it Clostridioides difficile} infected patients treated with
an experimental fecal microbiota transplantation.
Our statistical and topological data analysis are able to  detect  interesting  patterns  among patients and donors.
It also provides visualization of DNA sequences in the form of clusters and loops.
}
% =============================================
%
 %KEY WORDS AND PHRASES
%
% =============================================
\vspace{0.4in}

%%%% SECTION 1 %%%%%%%%%%
%%%%%%%%%%%%%%%%%%%%5
\section{Introduction}
Topological data analysis has become a formidable technique for analyzing high-dimensional data, especially when the
purpose is for classification and discrimination.  Methodological advancement has been rampant along with 
applications to medical or scientific data, see for example \cite{Nicolau2011}, \cite{heo2012} and more recently \cite{kovacev2014}.
In this paper we propose making use of computational topological techniques to analyze gut microbiome data at the basic
DNA sequence level based on data collected from sequencing the 16S rRNA gene.  We are particularly interested in seeing
changes in patient gut microbiome for a certain hypervirulent infectious disease following a radical experimental procedure
that is gaining widespread attention and usage in medicine. 

{\it Clostridioides} (formerly {\it Clostridium}) {\it difficile} (\emph{C.\ difficile}) infection (CDI) is the most frequent cause of healthcare-associated 
infections and its rates are growing in the community \cite{Kelly,Loo}.  
One of the major risk factors for developing CDI is through antibiotics. The 
healthy and diverse bacteria which reside within the colon are the major defense against the growth of {\it C.\ difficile}.  
Antibiotics kill these bacteria and allow {\it C.\ difficile} to multiply, produce 
toxins and cause disease. The current standard of care for this infection are the antibiotics: 
metronidazole, vancomycin and more recently, fidaxomicin.  The efficacy of these antibiotics is limited 
as vancomycin and metronidazole also suppress the growth of anaerobic bacteria such as {\it Bacteriodes 
fragilis} group which protect against proliferation of {\it C.\ difficile}.  The efficacy of the recent narrower spectrum fidaxomicin is
still under investigation although the initial data shows promise, \cite{Louie}.
The persistent disruption of healthy 
colonic flora may in part explain the reason for recurrences following a course of treatment with these 
antibiotics.     

As an alternative to antibiotic 
therapy for CDI, in particular for recurrent and refractory diseases, is to infuse healthy gut bacteria directly into the 
colon of infected patients to combat {\it C.\ difficile} by a procedure known as fecal microbiota transplantation (FMT).  
FMT is a process in which a healthy 
donor's 
stool is infused into an affected patient.  This can be performed using
a colonoscope, nasogastric tube, enema, or more recently, in capsulized pill form, \cite{khanna2016}.  FMT serves to reconstitute the altered colonic flora, in contrast to treatment with antibiotic(s), 
which can further disrupt the establishment of key microbes essential in preventing recurrent CDI. The 
literature reveals a cumulative clinical success rate of over 90\% in confirmed recurrent CDI cases \cite{Gough}.

There has been a growing interest into the microbiome of CDI patients \cite{Dethlefsen, Manges,Vincent,Petrof,Schubert2014}  especially those 
involved with FMTs \cite{Shahinas,van13,hamilton2013, song13, weingarden2014,Seekatz}.  In case of the latter, there are 
differences in:
the route of administration with all forms covered; different donor selection criteria, some used family members, some used a pool of donors; different
sample sizes, although all studies had small sample sizes; and different sequencing procedures and equipment.  Despite these differences, there seems to be two fundamental 
points of agreement across all studies.  The first is that CDI patients have low diversity in their microbiome, and that after receiving an FMT(s), their diversity was increased. The second fundamental agreement is CDI patients who were treated with FMT undergo changes in their microbiome that 
at least initially have similarities to that of their donors.  
This paper provides further reinforcing evidence to support these two fundamental points in the framework
of FMT delivered by enema, with a small exclusive donor pool.  The novelty comes from using computational
topological techniques to demonstrate this.

We now summarize the paper.  In Section \ref{sec:CDI} we explain the details of the clinical data.
In Section \ref{sec:prelim} we provide topological preliminaries where we go over
the Vietoris-Rips complex, and three topological descriptors: barcodes, {persistence diagrams}, and {persistence landscapes}.
In the following Section \ref{sec:avgPL}, we apply these techniques and demonstrate
the added value of the topological approach at the basic DNA sequence level.  
We complete our article in Section \ref{sec:conclusion} with a summary of  the key findings. 

\section{CDI FMT and 16S rRNA data} \label{sec:CDI}

An earlier report provided details with regard to donor screening and the protocol used \cite{kassam12}.
Including the earlier study, FMT was administered to 94 patients and several patients required multiple FMTs to achieve resolution \cite{lee2014}.  
From this patient pool we selected 19 patients, not necessarily randomly, for sequencing their 16S rRNA gene prior to treatment, pre-FMT, followed by a
post treatment, post-FMT.  
Consequently, the data that is about to be presented
should not be interpreted with respect to the efficacy of FMT, as the purpose is to better understand the microbial changes that came about as a result of
an FMT(s).

The gender of the patients was 63\% female.  The average age (to the time of their first FMT) was 77.11 with a standard deviation of 9.54 years,
range 49 to 92 years. 
In-hospital patients accounted for 53\% while the total peripheral white blood count ($\times 10^{9}$/L) had a mean and standard deviation of 14.6 and 10.87, respectively.  Three patients had temperature greater than $38^o$ C, and 11 patients experienced abdominal pain.  Approximately half (53\%) of the patients were on proton-pump inhibitors.  
Eight patients were refractory to treatment with metronidazole and four patients were refractory to
vancomycin. These four patients were also refractory to metronidazole.  No patients were on fidaxomicin as it had not been available to
the public at that time.
The patient characteristics are provided in Table \ref{table1}, below.

\begin{table}[h]%\small
\begin{center}
\begin{tabular}{|l|r|}\hline
Covariate& \\\hline
Age--years& \\
\hspace{.5 cm}Mean $\pm$ Standard Deviation&$77.11\pm 9.54$\\
\hspace{.5 cm}Range &49--92\\
White Blood Count--$\times 10^9$ per litre& \\
\hspace{.5 cm}Mean $\pm$ Standard Deviation&$14.6\pm 10.87$\\
\hspace{.5 cm}Range&5--40\\\hline
Female--count(\%)&12 (63\%)\\
In-hospital--count(\%)&10 (53\%)\\
Fever--count(\%)&3 (16\%)\\
Abdominal Pain--count(\%)&11 (58\%)\\
Proton pump inhibitor--count(\%)&10 (53\%)\\
Refractory to Metronidazole(\%)&8 (42\%)\\
Refractory to Vancomycin(\%)&4 (21\%)\\
\hline
\end{tabular}
\caption{Patient demographics and pre-FMT conditions.}\label{table1}
\end{center}
\end{table}

Predisposing conditions that may have resulted in CDI were:  cellulitis, extreme fatigue, respiratory tract infections, septicemia, surgery and open wounds,  and urinary tract infections.  Some conditions were unknown. 
The majority of patients received the following antibiotics prior to contracting CDI: amoxicillin, azithromycin, cefazolin, cefprozil, cephalexin, ciprofloxacin, clindamycin, clarithromycin, cloxacillin, levofloxacin, moxifloxacin and nitrofurantoin.
Some patients claimed no prior antibiotics used.  In an interesting paper, the affects of  ciprofloxacin was studied \cite{Dethlefsen} using 16S rRNA deep sequencing.

The patients all had varying pre-FMT regimens to treat their CDI. All had undergone multiple rounds of traditional antibiotic therapy with metronidazole and/or vancomycin before being administered an FMT. Seventeen received at least one course of metronidazole monotherapy, 18 received vancomycin monotherapy, 6 received vancomycin taper, and 3 received concomitant metronidazole and vancomycin therapy. Patients generally received two courses of metronidazole followed by multiple courses of vancomycin before receiving FMT(s). 
The number of days every patient received each therapy are summarized by their means and standard deviations
and reported below in Table \ref{table2}.

\begin{table}[h!]%\small
\begin{center}
\begin{tabular}{|l|r|}\hline
Standard of Care Treatment& \\\hline
Metronidazole(\%)&17 (90\%)\\
Metronidazole--days& \\
\hspace{.5 cm}Mean $\pm$ Standard Deviation&23.74$\pm$ 16.51\\
\hspace{.5 cm}Range &01--70\\
Vancomycin(\%)&18 (95\%)\\
Vancomycin--days& \\
\hspace{.5 cm}Mean $\pm$ Standard Deviation&32.58$\pm$ 20.94\\
\hspace{.5 cm}Range &7--86\\
Vancomycin Taper(\%)&6 (32\%)\\
Vancomycin Taper--days& \\
\hspace{.5 cm}Mean $\pm$ Standard Deviation&37.58$\pm$ 93.86\\
\hspace{.5 cm}Range &38--390\\
Metronidazole-Vancomycin(\%)&3 (16\%)\\
Metronidazole Vancomycin--days& \\
\hspace{.5 cm}Mean $\pm$ Standard Deviation&5.21$\pm$ 18.55\\
\hspace{.5 cm}Range &2--80\\
\hline
\end{tabular}
\caption{Pre-FMT treatment for CDI.}\label{table2}
\end{center}
\end{table}

Each patient had two stool samples sequenced, one representing a pre-FMT sequence, and one representing a post-FMT
sequence.
We attempted to sequence each patient prior to them receiving any FMT which for the most part occurred except for one
patient who failed their first FMT and their stool sample sequenced was taken right after that event but prior to their next FMT.
This patient had multiple FMT failures, hence we took that stool sample as the pre-FMT sequence.  We also took a stool sample 
following each patient's pre-FMT sample with at least one FMT in between.  For the most part, the latter occurred following their
last FMT which was the case if the patient resolved their CDI, but if they did not, then the post-FMT sequence was not
necessarily their last.  We would also like to make clear that 4+-FMT means that a patient had at least four treatments but could
have had more.  We indicate such as `4+' to be consistent with our clinical paper, \cite{lee2014}.

The breakdown of the data is presented in Table \ref{tab:3} with the above qualifications.  All patients received
a single
treatment, 1-FMT and 9 of them clinically resolved their CDI.  All who failed the first treatment went on
to receive a second treatment.  Of the remaining patients who received 2-FMT, 2 resolved, while the
remaining went onto to receive a third treatment.  There were 2 successes, and 1 failure, meaning they
did not go on to receive additional treatments.
Of the remaining patients who went on to $4+$-FMT, 1 resolved, 1 resolved with antibiotics used in between
treatments, and there were 3 failures.

\begin{table}[h]%\small
\label{table3}
\begin{center}
\begin{tabular}{|c|c|c|c|c|}\hline
FMT&Resolution&Resolution*&Failures&Total\\\hline
1&9&-&0&9\\
2&2&-&0&2\\
3&2&-&1&3\\
$4+$&1&1&3&5\\\hline
-&14&1&4&19\\\hline
\end{tabular}
\caption{Clinical resolution of CDI following FMT(s).  Patients who received antibiotics in-between FMTs are marked with
an asterisk.}
\label{tab:3}
\end{center}
\end{table}

There were several patients who received antibiotics in between FMTs as described in our previous report \cite{lee2014}.
In addition four healthy volunteers served as donors and were screened for transmissible pathogens as was outlined 
in an earlier report \cite{kassam12}. The donors took no antibiotics for 6 months prior to stool donation.  Seven donor samples 
taken at various times were sequenced. 

All \textit{C.\ difficile} infections were confirmed by in-hospital real-time polymerase chain reaction (PCR) testing for the toxin B gene.
This study sequenced the forward V3-V5 region of the 16S rRNA gene from 19 CDI patients who were treated with FMT(s).  A pre-FMT, a corresponding
post-FMT, and 7 samples from four donors, corresponding altogether to 45 fecal samples were sequenced.
All sequencing was performed on the 454 Life Sciences, GS Junior Titanium Series.
The Qiagen Stool Extraction Kit  (Omega BIO-TEK, Norcross, Georgia) was used to extract the DNA from the fecal samples following the `stool DNA protocol for pathogen detection'.
Subsequent DNA amplification was done using PCR forward and reverse primers.
One round of DNA amplicon purification was performed using the QIAquick PCR Purification Kit (Qiagen, Valencia, CA) followed by two rounds of purification using Agencourt AMPure XP beads (Beckman Coulter Inc., Mississauga, ON).  

We examined
19 pairs of pre-FMT and post-FMT patients, as well as donors, selected from the 94 CDI patients treated by the first author over the period 2008-12, \cite{lee2014}. 
The selection was not random and was chosen to reflect as wide a variability as possible.  Furthermore, we deliberately chose 4 FMT failures to try and
better understand what FMT can and cannot do.  It was perhaps through the failures that we learned the most.  The failed FMT cases were ultimately
resolved with antibiotics even though some patients were refractory to metronidazle and/or vancomycin beforehand.  
Indeed FMT acted as a `gut primer' for antibiotics to fulfill it's role in clearing recurrent and refractory CDI. 
Although this phenomenon of
failed FMT patients resolved with antibiotics afterwards has been observed in our work, \cite{lee2014}, as well as others, see for example \cite{rubin13}, as far as we are
aware this is the first paper that has pursued this from a microbiome point of view.
The data comes from the clinical work of the fourth author who is a practising physician.
Some of this data
was also examined in \cite{martinez2016} and \cite{rush2016}.

Table \ref{tab:desc_total} below, provides some descriptive statistics about the total and unique number of DNA sequences found in the 45 samples. 
Note that it is impossible to ensure that we have an approximately equal number of sequences in each sample \cite{li2008}. 

\begin{table}%[position specifier]
  \centering
  \begin{tabular}{| c | c | c | c | c | c | c |}
  \hline
  & min   & max    & mean        & median  & S.D.\\ \hline
  Pre-FMT          & 147 (3230) & 879 (15140)    & 428.89 (9534.9)    &  364 (9777 )   & 217.26 (3545.9) \\ \hline
  Post-FMT        & 185 (2294)  & 1114 (28566)  & 486.42 (11308.11) &  460 (10570)   & 230.57 (6642.6)  \\ \hline
  \end{tabular}
\vskip0.4cm
  \caption{Descriptive statistics for unique (total)  number of DNA sequences of pre-FMT and post-FMT samples.}
  \label{tab:desc_total}
\end{table}

Consider the DNA sequences $\lambda$ and $\mu$. Let $x$ be the number of point dissimilarities between DNA bases in $\lambda$ and $\mu$, and let $x_\lambda$ and $x_\mu$ be the number of DNA bases in $\lambda$ and $\mu$, respectively. Let $x_O$ be the number of places where one sequence contains a DNA base and the other an $O$, a gap. Let $y_\lambda$ and $y_\mu$ be the number of gaps in $\lambda$ and $\mu$, respectively. Let $z$ be the length of the alignment sequences; we can assume they are
equal otherwise we choose the smaller.  The various dissimilarity metrics are the following.

The {\it one-gap} distance $d_O$ is defined by 
\[d_O(\lambda,\mu)=\frac{x+\min\{y_\lambda,y_\mu\}}{\min\{x_\lambda+y_\lambda,x_\mu+y_\mu\}}.\]
It treats a string of $O$'s flanked by any DNA bases in one sequence and the corresponding region in the other as one mismatch.

The {\it no-gap} distance $d_N$ is defined by 
\[d_N(\lambda,\mu)=\frac{x}{\min\{x_\lambda,x_\mu\}}.\]
It ignores gaps in a sequence and the corresponding region in the other sequence.

The {\it each-gap} distance $d_E$ is defined by 
\[d_E(\lambda,\mu)=\frac{x+x_O}{z}.\]
It treats every DNA base-placeholder pair as a mismatch.

These metrics produce true mathematical distance matrices, which are symmetric and contain zeros on the diagonal.
For this and most studies, the metric of interest is the one-gap metric, \cite{rush2012}.  As an example, suppose there are two DNA sequences with the following base pair orientation:
\texttt{ATGCATGCATGC} and  \texttt{ACGC---CATCC}.
Here there are two mismatches and one gap. The distance is calculated as the number of mismatches divided by length of the shorter sequence. The length of the shorter sequence is 10 base pairs, since the gap is considered a single position. Hence distance is 0.3, \cite{rush2012}.
We can use the other definitions of distance, but the definition provided is the one most commonly used. In addition, the results do not vary significantly. 
Most of the DNA sequences found in the samples appear several times, and hence the distance between them will be zero as they are identical. For this reason, the unique DNA sequences are taken.

Table \ref{tab:desc_total} shows the minimum number of unique sequences is 147. DNA sequencing does not provide exact results, hence as the number of sequences in a sample increases, some mutations inevitably occur and these are recorded as unique sequences \cite{rush2012}. In other words, as the total number of sequences increases, the number of unique sequences is also inflated. For this reason, it is necessary to subsample from the number of unique sequences. The smallest number of unique sequences is 147, hence a weighted subsample of size 147 is taken from the number of unique sequences for our research. 
The pairwise distance between the 147 sequences is calculated in each of the 45 samples using the one-gap metric.
Thus the data used for the primary analysis is 147 $\times$ 147 distance matrices for each of the 45 patients and donors.

%%%% SECTION 3 %%%%%%%%%%
%%%%%%%%%%%%%%%%%%%%5

\section{Some Topological Preliminaries}\label{sec:prelim}

Persistent homology is a branch of computational topology that has been popularized by \cite{Edels2002} and 
\cite{ZomCar2005}.
Several researchers have shown that persistent  homology  works well on detecting topological and geometrical  features in high dimensional data, see \cite{edelsbrunner2008} and  \cite{Nicolau2011},  for example. % and \cite{fasy2013}, name a few.
Let us suppose that we have points on a manifold whose dimension is not necessarily known. 
Topology   studies the connectivity of these points.
Each point  is replaced by a disk~(ball) with  radius $\eps$ centered at each point.  
If the disks overlap, connect  those  center points with edges.  As the radius increases, more points will be connected and so the number of connected components will decrease. This is analogous to clustering analysis in statistics. The connected components~(clusters) are considered as 0-degree topological features.
The number of connected components  is denoted as the 0-th Betti number, $\beta_0.$
Connecting the points with edges will also create simplices (convex hull of a geometrically independent set of points) and  produce topological features in higher dimensions. The low dimensional simplices are well known;   a vertex (0-degree simplex), an edge (1-degree simplex), a triangle (2-degree simplex), and a tetrahedron (3-degree simplex).
A loop  is also called a 1-degree  topological feature  and a void is a 2-degree topological feature.
The $k-$th Betti number $\beta_k$ counts the number of $k-$degree topological features ($k-$dimensional `holes').

\begin{figure}[htp]
  \begin{center}
\begin{tabular}{ccc}
 \includegraphics[scale=0.25] {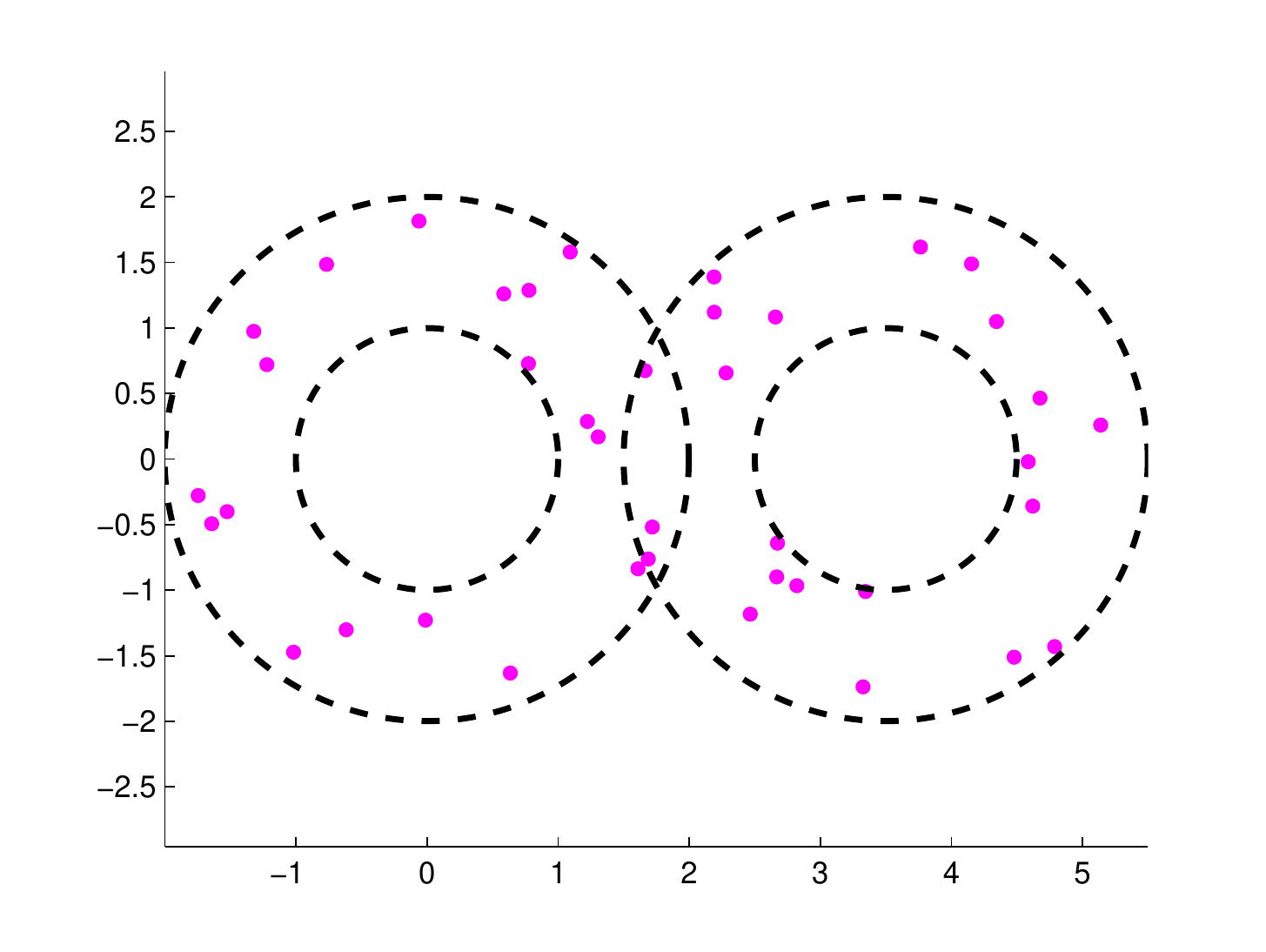} & \includegraphics[scale=0.25] {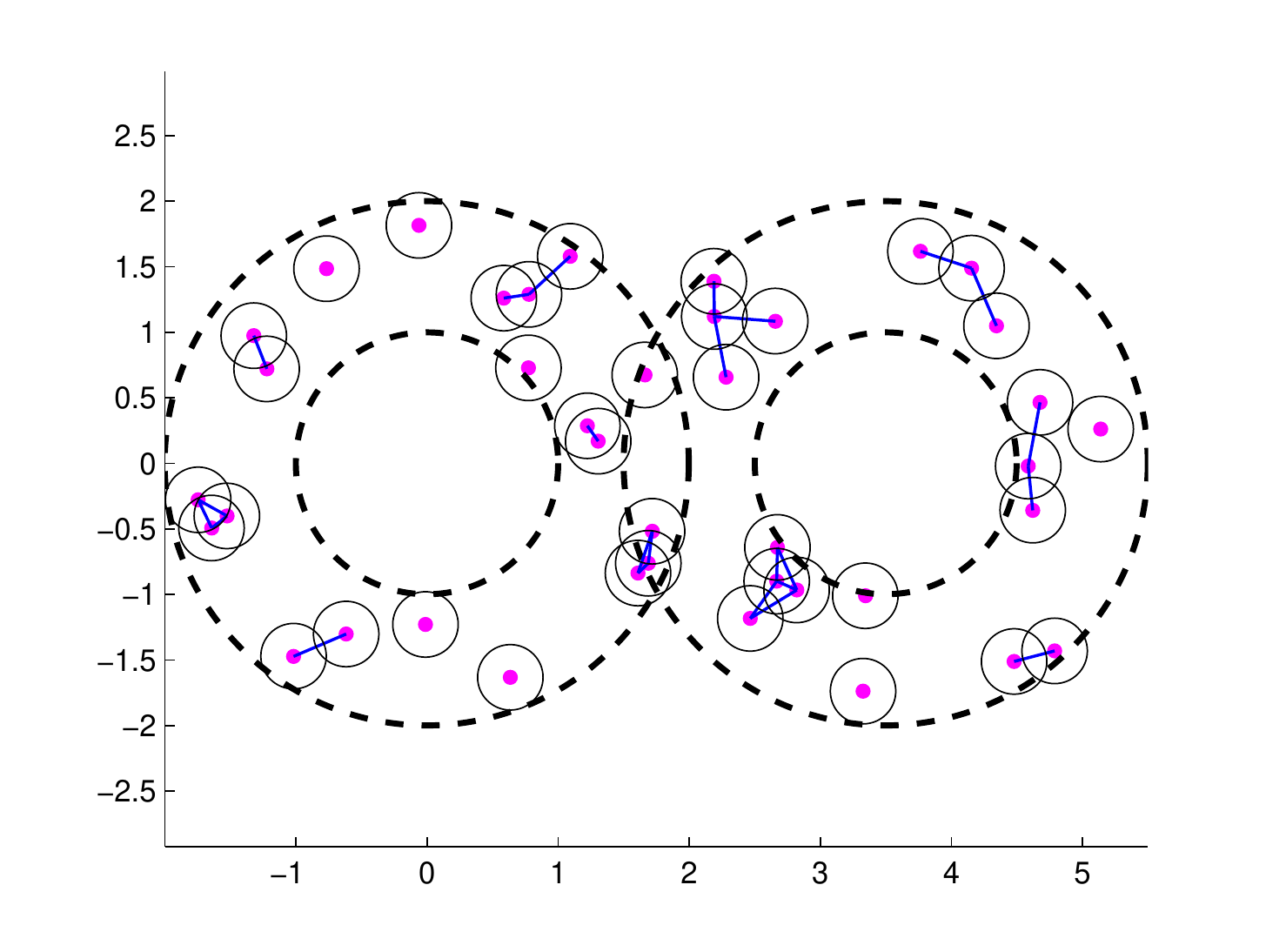} & \includegraphics[scale=0.25] {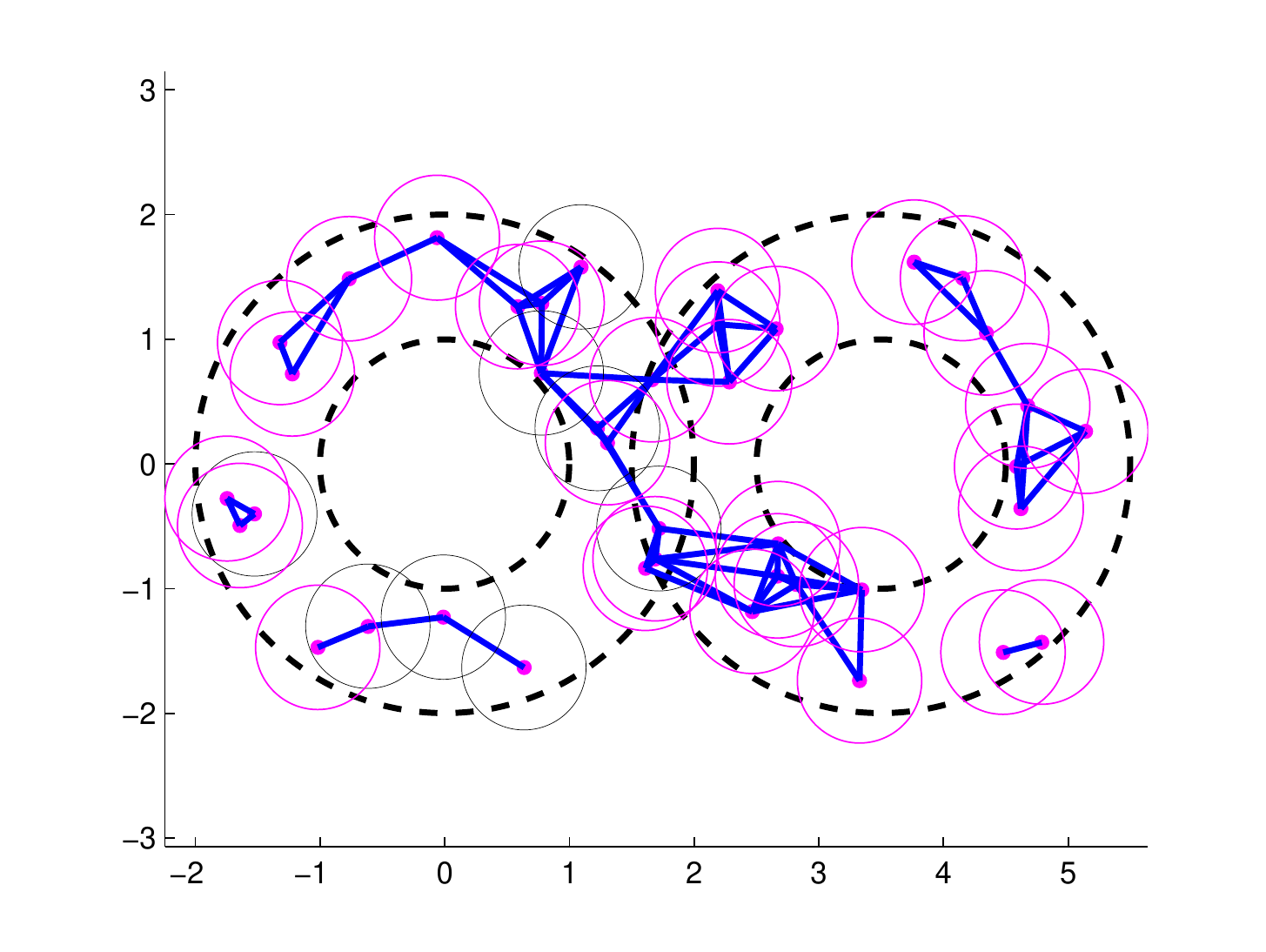}  \\
\includegraphics[scale=0.25] {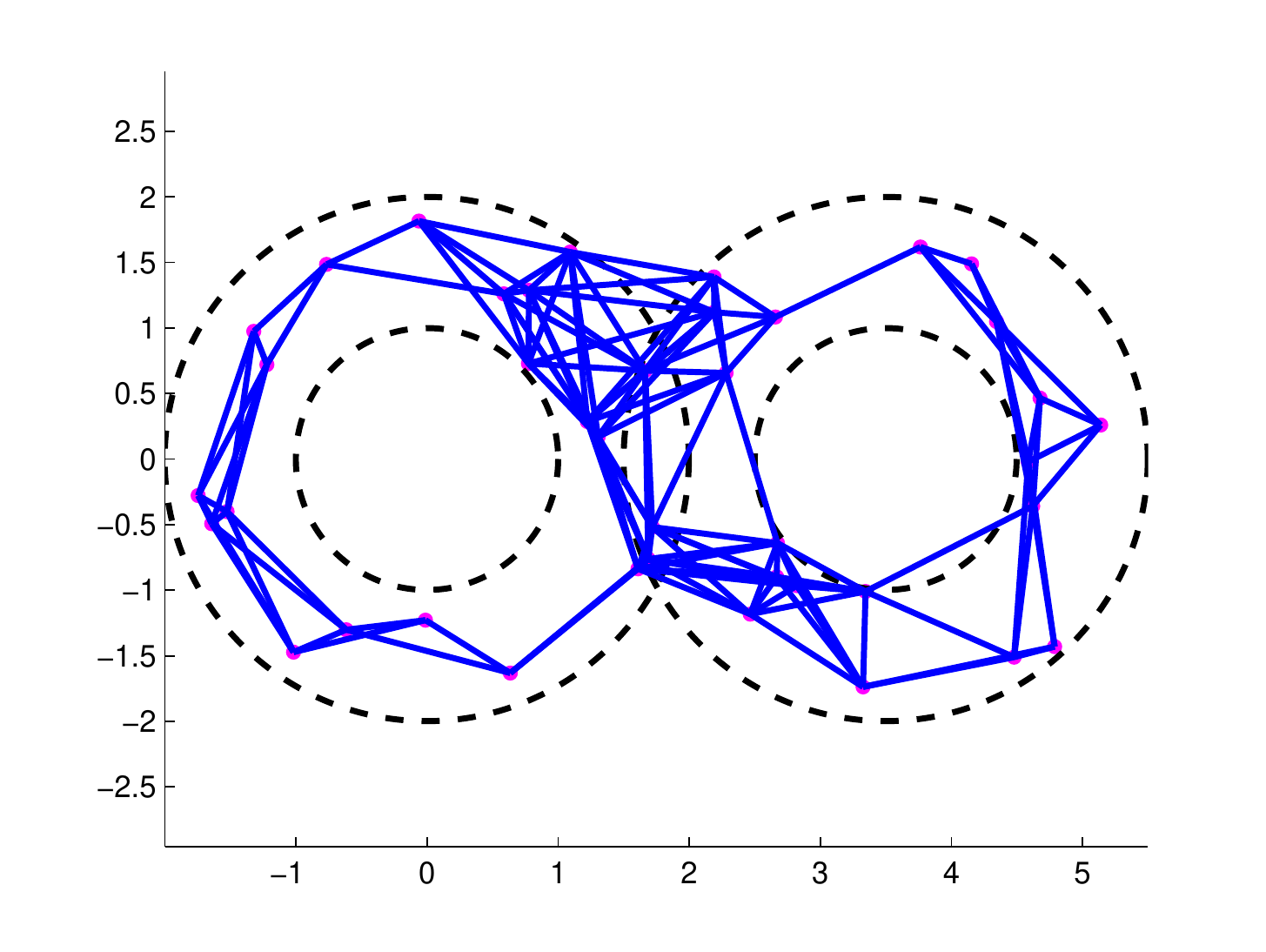} & \includegraphics[scale=0.25] {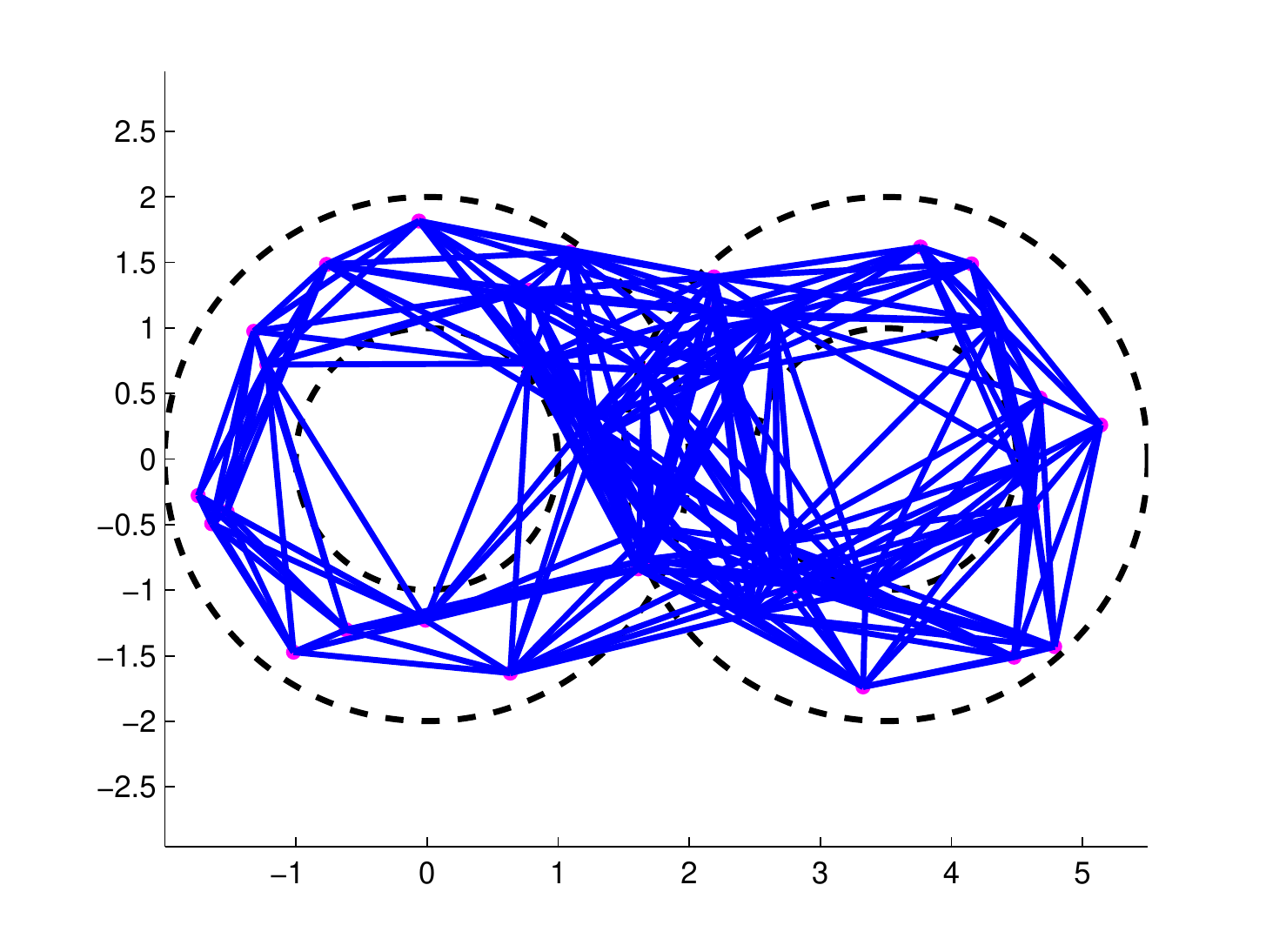} & \includegraphics[scale=0.25] {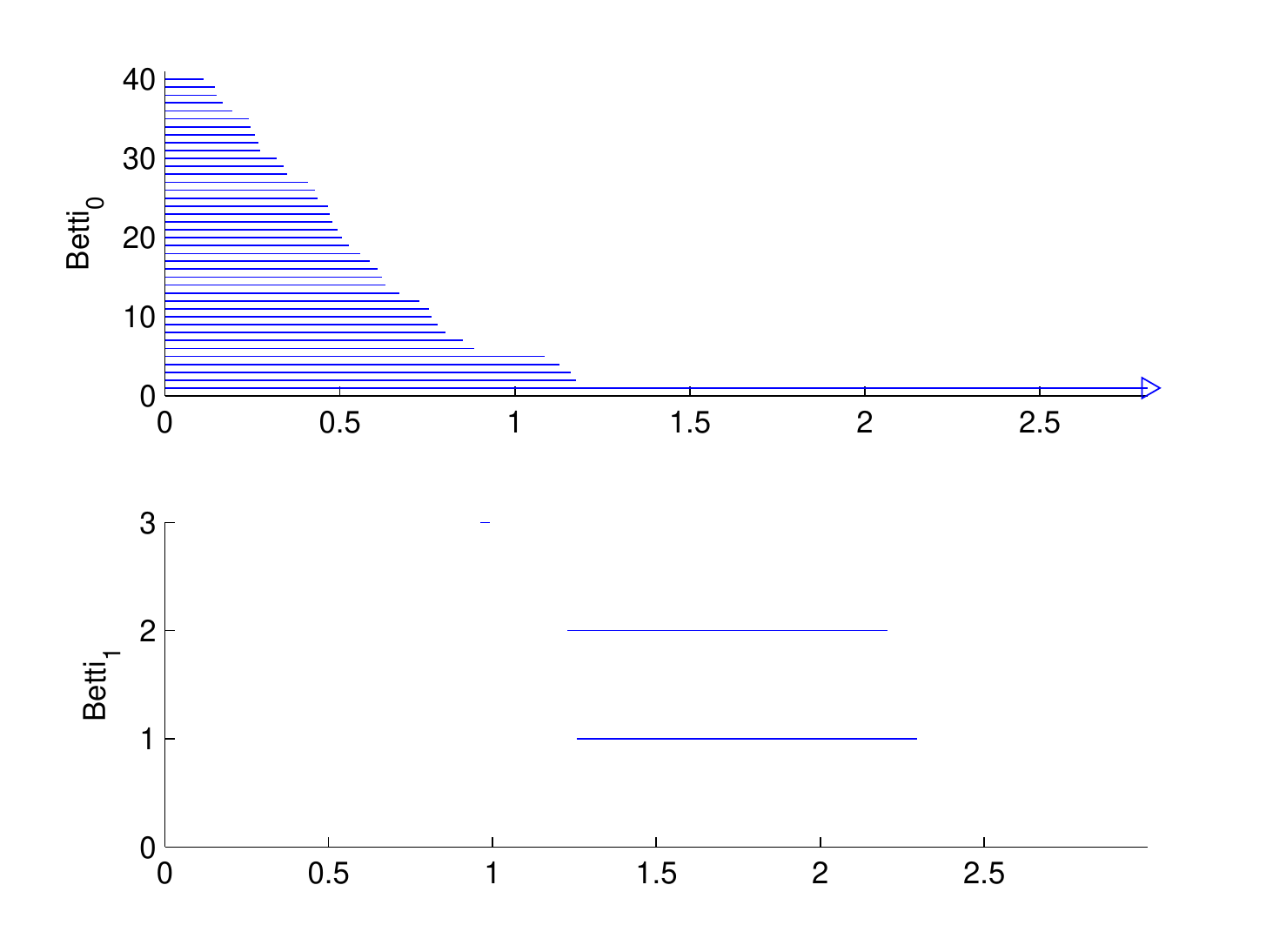} 
\end{tabular}
  \end{center}
  \caption{Forty points are sampled randomly from a double annulus with inner circle radius 1 and outer circle radius 2.  
Forty components (top left) yield $\beta_0 = 40$ at $ \eps= 0$;  nineteen (top middle) connected
components, that is,  $\beta_0 = 19$ at  $\eps= 0.5$; top right $\beta_0 = 5$ at
$\eps= 1.0$;  the two most persistent loops (bottom left and middle) in the middle of the
double annulus that are born at about  $\eps=1.5$  die around $\eps= 2.4$
showing two `true' persistent loops. The
long $(0, \infty)$ bar in $\beta_0$ (bottom right) barcode indicates the one persistent component while two longer bars in $\beta_1$  barcode indicate two loops.}
  \label{fig:doubleAnn}
\end{figure}

Persistent homology studies the history of topological features as the parameter $\varepsilon$ varies.
It records  the time at which a topological feature appears and disappears.
The birth, death, and  survival time of features are recorded as a {barcode} \cite{CZCG2004a}.
A {true} feature in the data  lives over a long time while noise is short lived.
We illustrate a fundamental idea of persistent homology  in Figure~\ref{fig:doubleAnn} with 40 randomly selected points from a double  annulus.  The loops in the  middle of each annulus  are prominent  1-degree features  and their  intervals in the barcode shows their persistence.

%%% FIGURE 1%%%%%%%%%%%

\begin{figure}[htp]
  \begin{center}
\begin{tabular}{cccc}
\includegraphics[scale=0.1200] {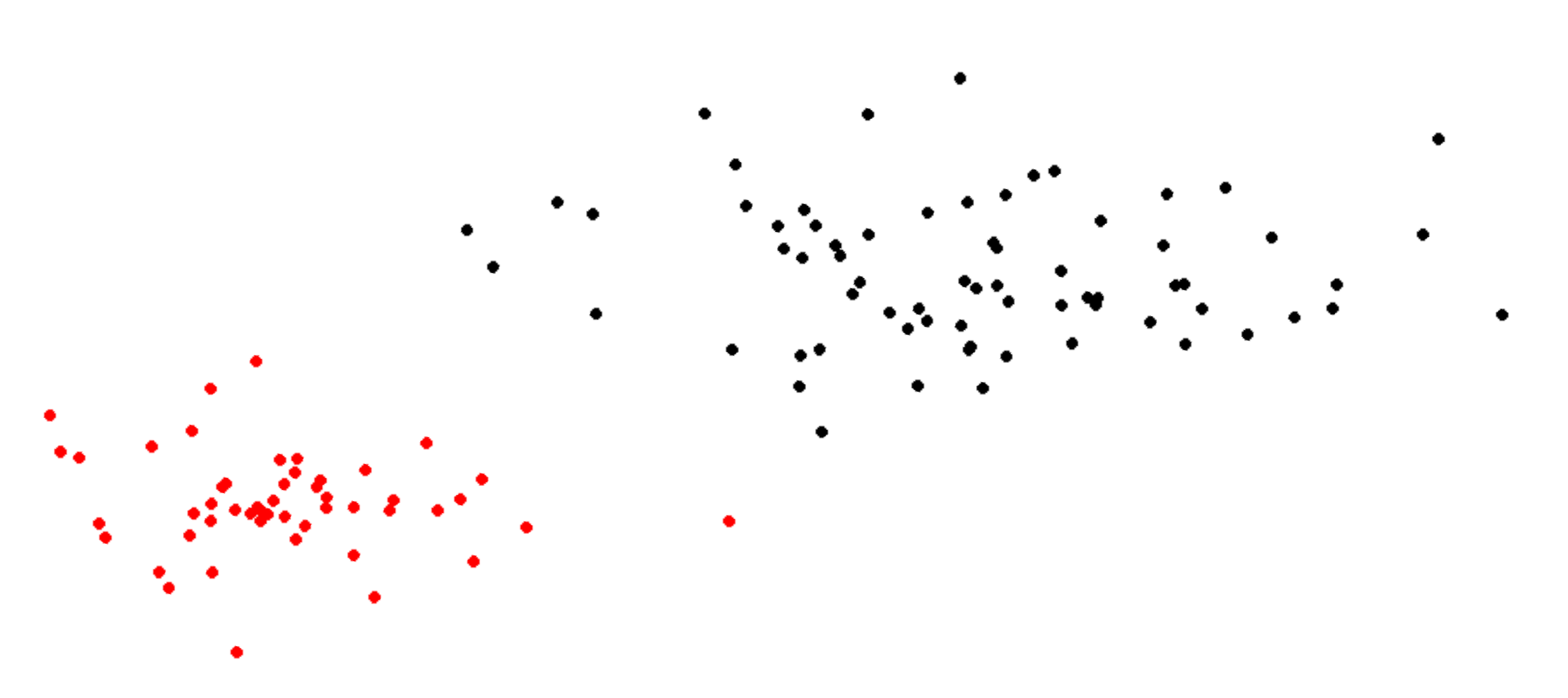}  & \includegraphics[scale=0.10] {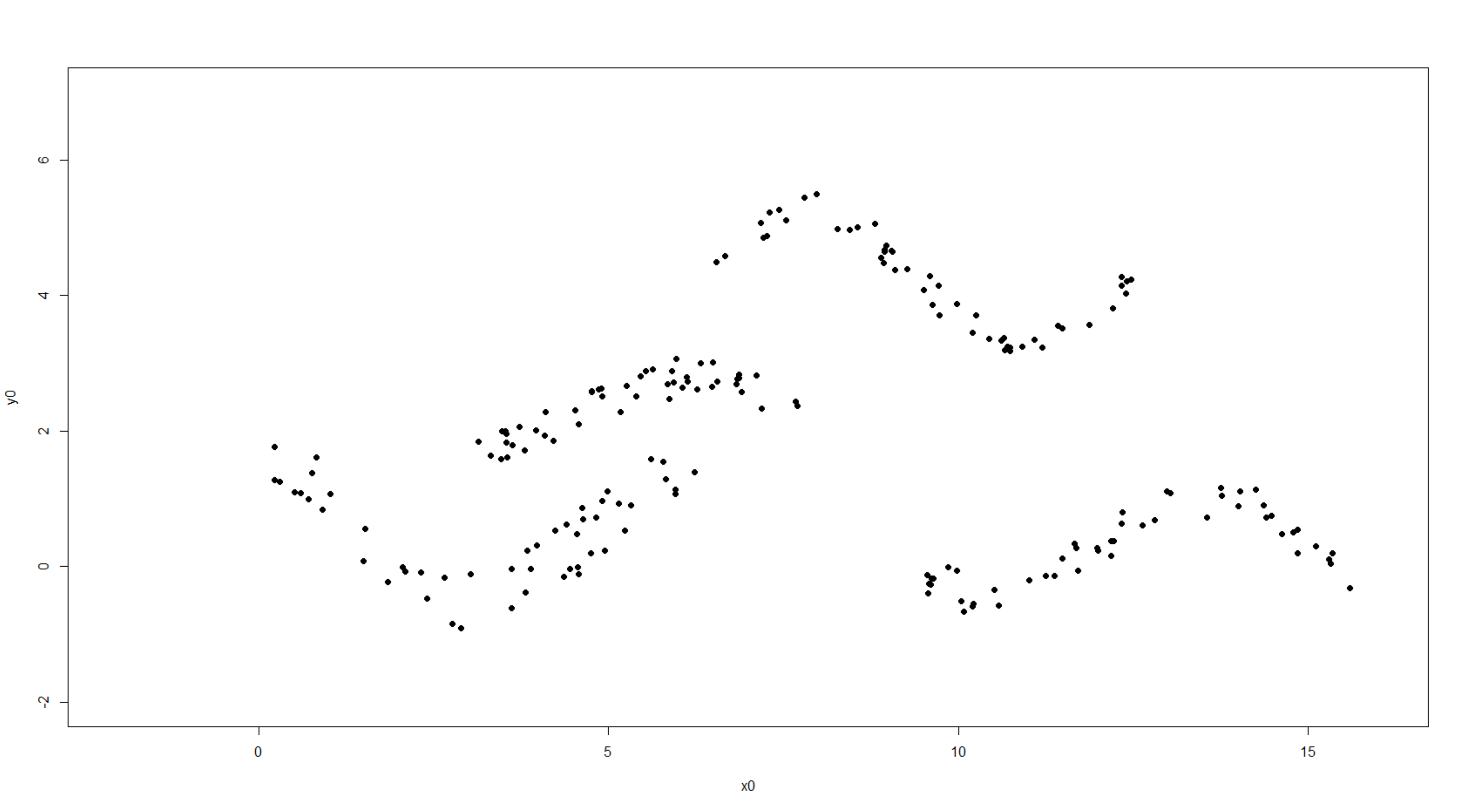}
& \includegraphics[scale=0.20] {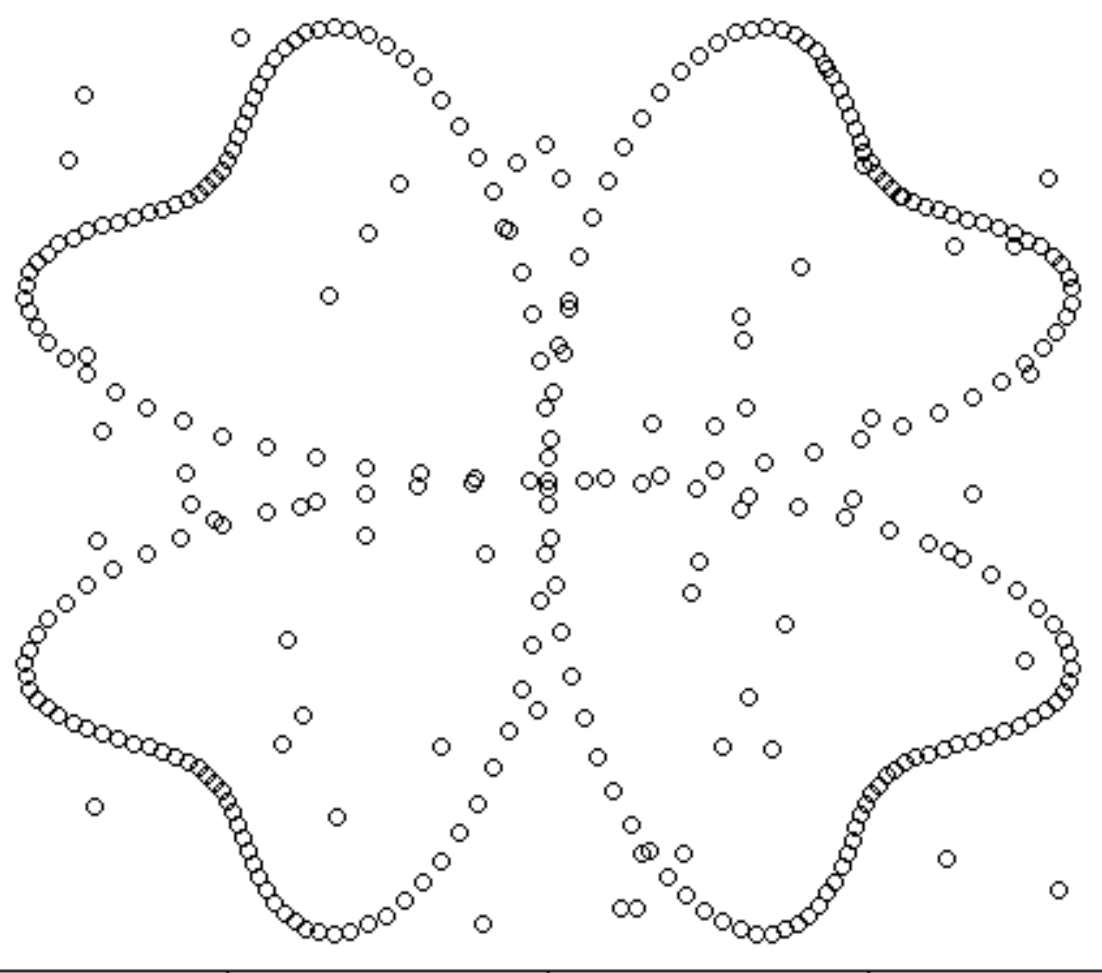} & \includegraphics[scale=0.25] {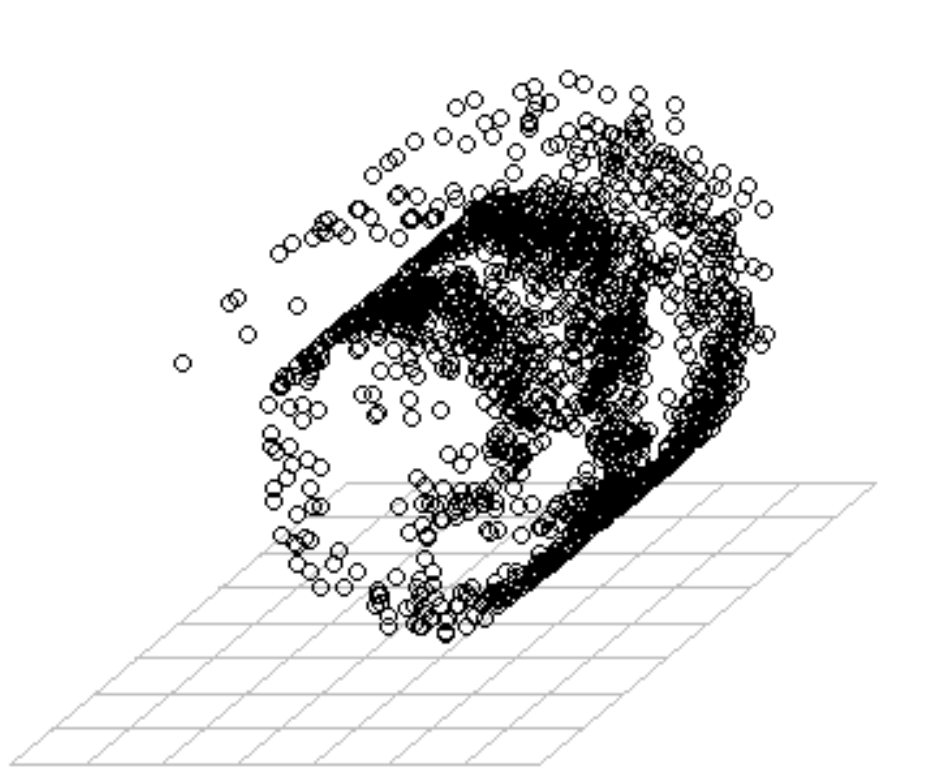}\\
\end{tabular}
  \end{center}
  \caption{Synthetic examples of clusters in different forms. From `blobs'-intuitively  familiar clusters, to curves, `leaves', and Swiss roll.}
\label{fig:syndata}

\end{figure}

\begin{figure}[htp]
  \begin{center}
\begin{tabular}{ccc}
\includegraphics[scale=0.15] {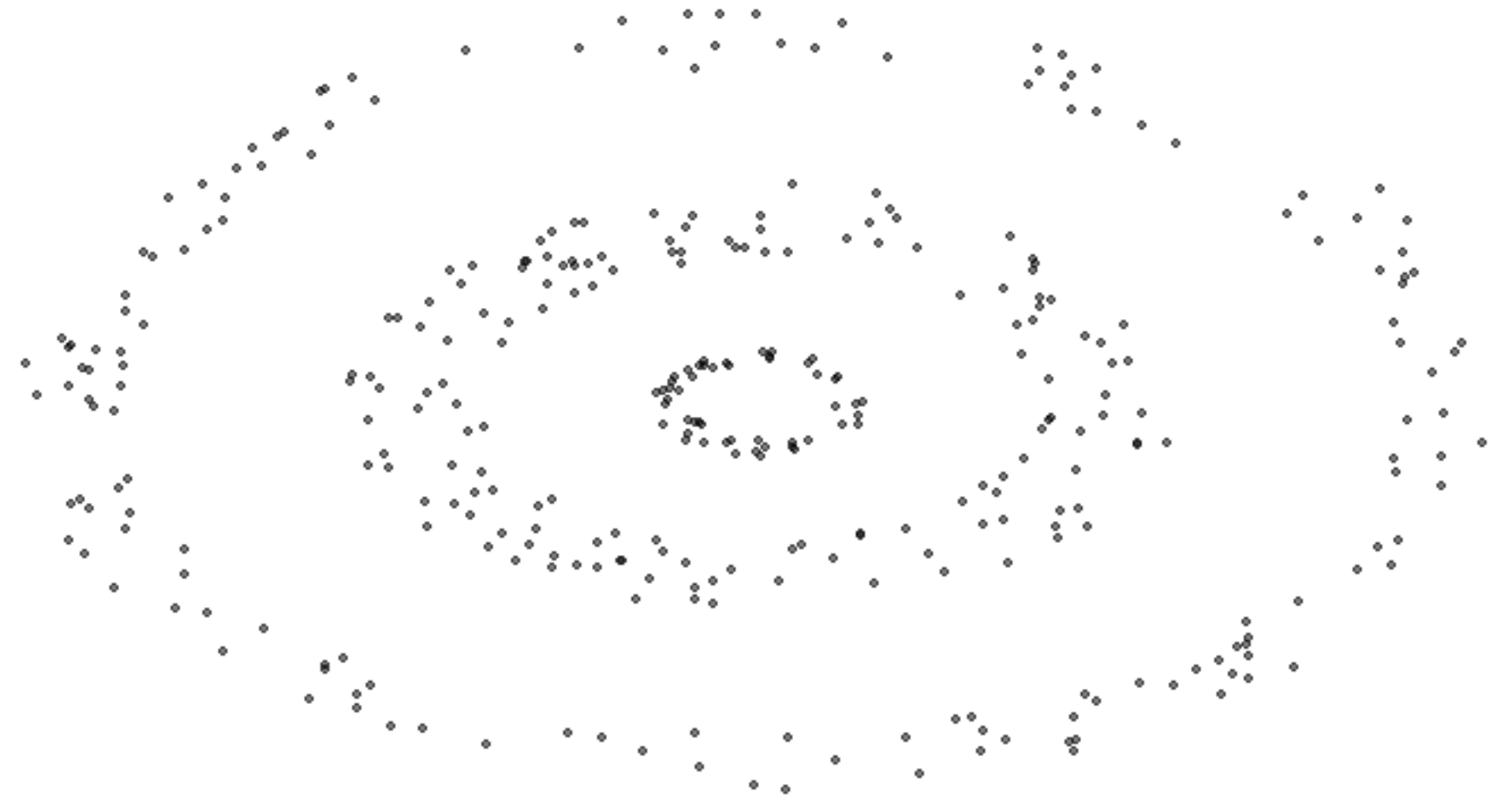}  & \includegraphics[scale=0.15] {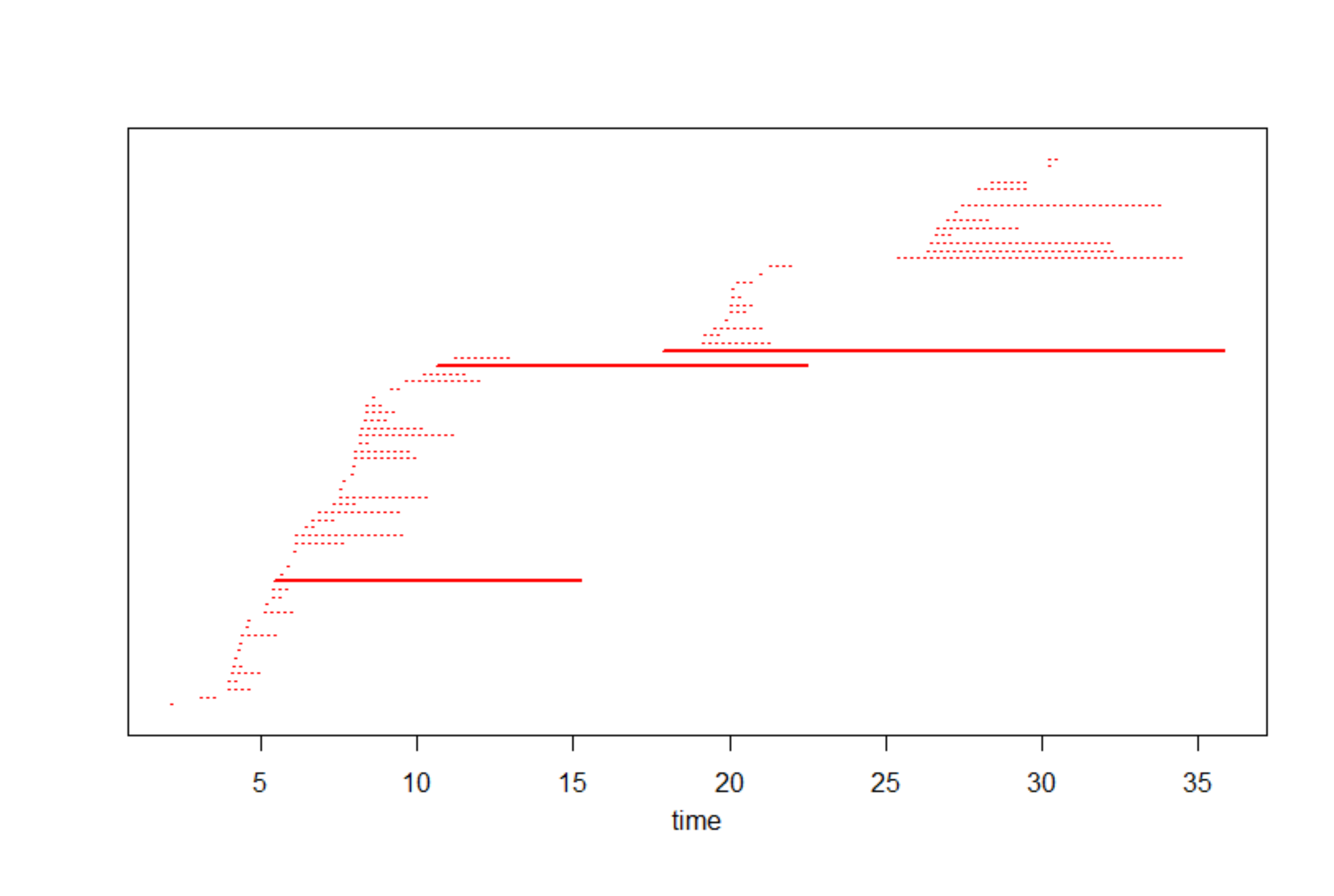} & \includegraphics[scale=0.15] {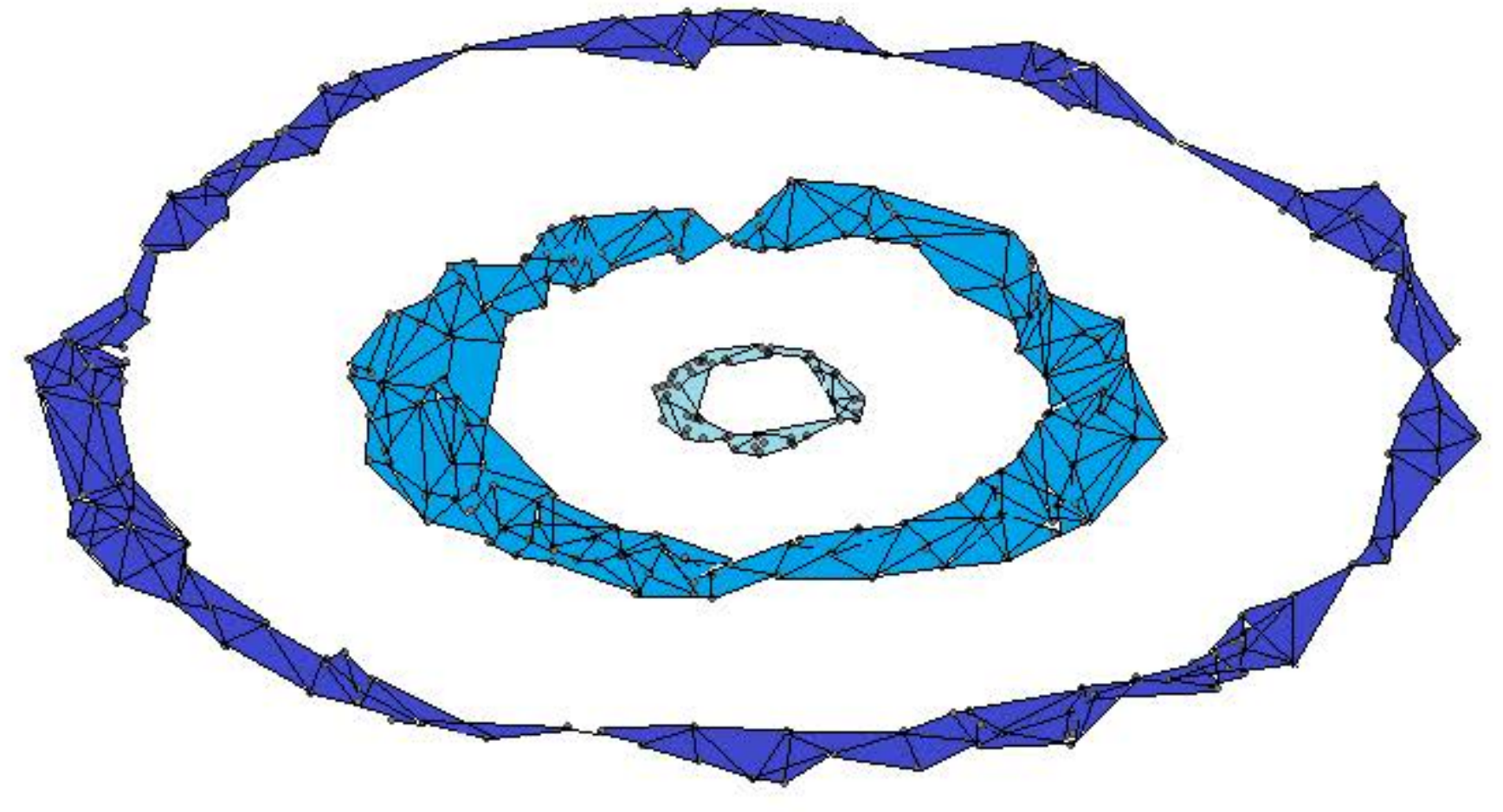}\\
\includegraphics[scale=0.10] {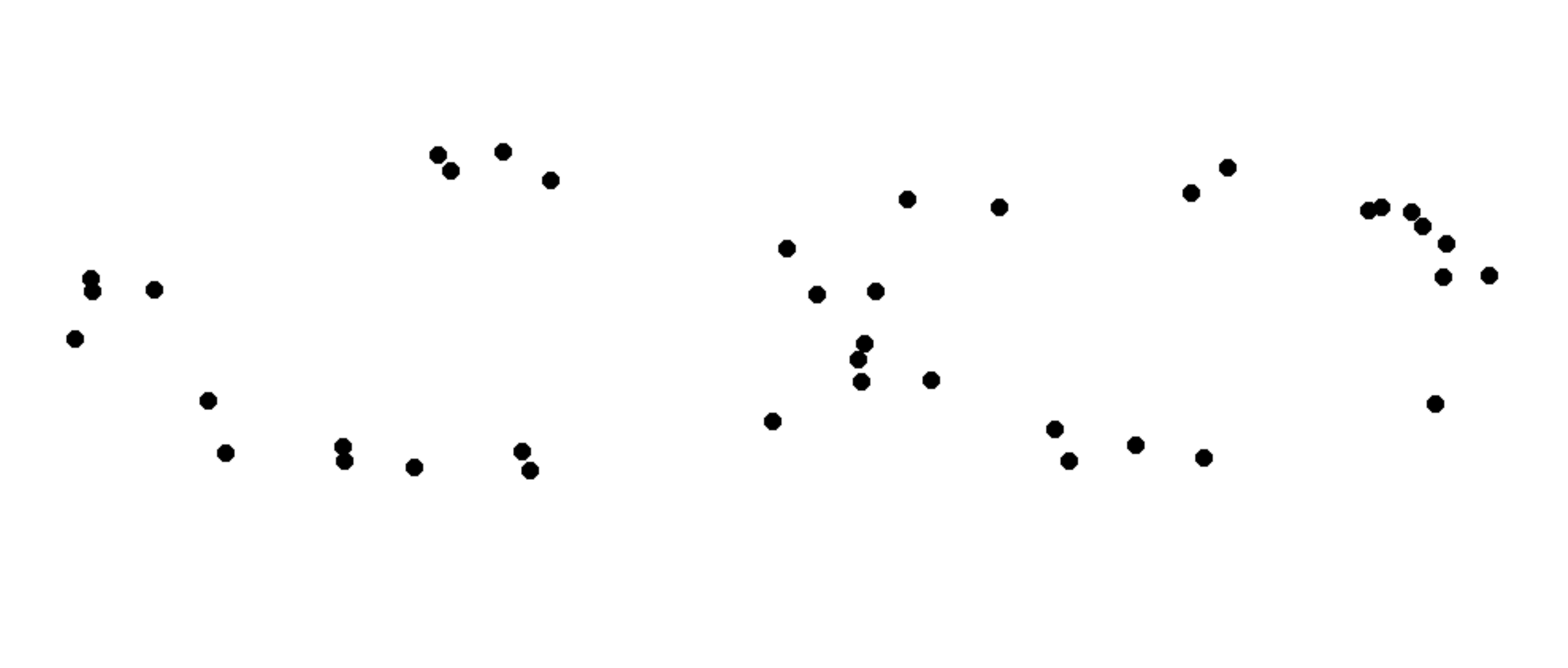}  & \includegraphics[scale=0.10] {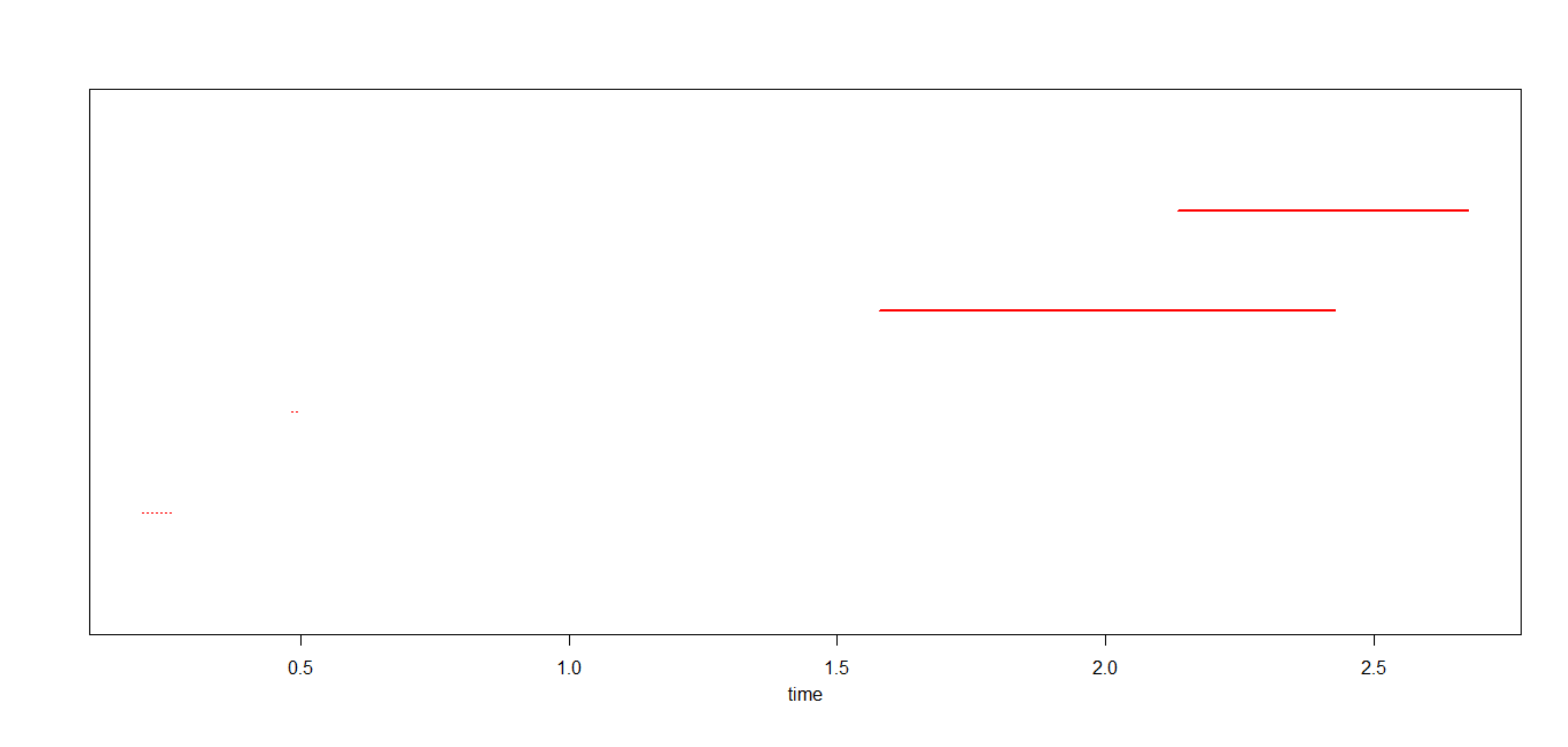} & \includegraphics[scale=0.15] {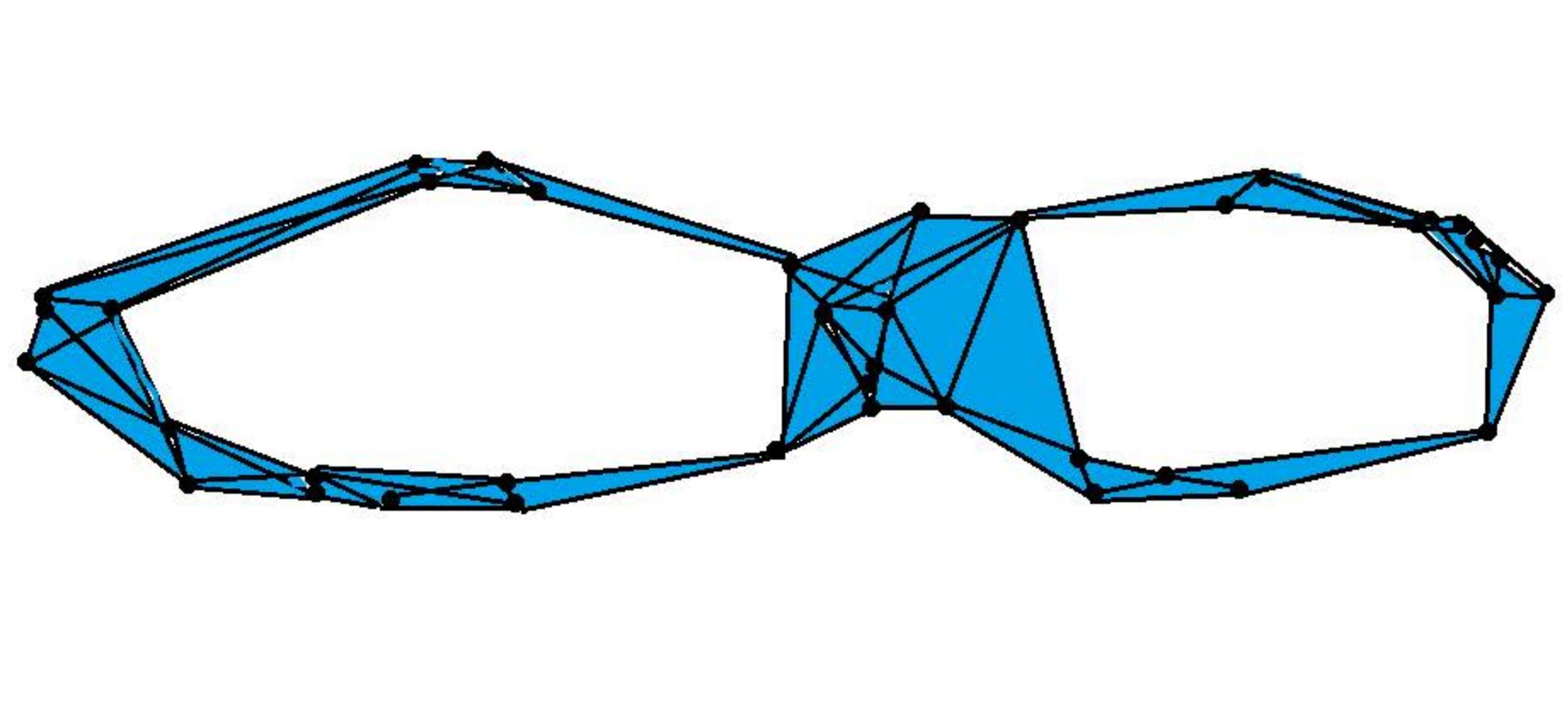}
\end{tabular}
  \end{center}
  \caption{Top: random points are generated to mimic concentric rings analysed in \cite{benhur2001}. The three long intervals in the barcode shows three persistent loops.   The three persistent loops are drawn which correspond to three clusters in \cite{benhur2001}. Bottom: point clouds of numeric number 8 with noise.  $\beta_1$-barcode indicates two persistent signals. Two possible clusters are detected as two loops, that is, 1-degree topological feature. }
\label{fig:rings}
\end{figure}

{Clusters  come in different shapes, see a few synthetic examples in Figure \ref{fig:syndata}.
Clusters are  homogeneous subgroups where the meaning of homogeneity is dependent on the types of similarity measure. 
In the double annulus, Figure \ref{fig:doubleAnn}, if we consider geodesic distance between points,  those points are connected to form simplices (edges and triangles).  All the edges and triangles together form a band which is homotopy equivalent to a circle (1-degree topological feature).  The authors in \cite{benhur2001} applied support vector clustering to concentric rings and were able to detect three rings as clusters. In Figure \ref{fig:rings}, we generate similar concentric rings as in \cite{benhur2001}.  Persistent homology analysis shows three persistent loops which correspond to three clusters in \cite{benhur2001}. Consider another example,  where points are randomly sampled from  the number 8 with noise, see Figure \ref{fig:rings}. 
What are the clusters in this data?  If one thinks of `blobs' as clusters, there are several patches made up with a few points.
However,  each loop, 1-degree topological feature,  is represented as the set of points which are {close} in the sense of the shortest path distance.}


To study the connectivity  of a space,  the space need not  be  represented  as a point cloud. 
All we need to know is how close  the points are in a space where they live.  
Thus point clouds or matrices of (dis-)similarity measurements among other data forms are input data for analysis of persistent homology.  
A good reference for algebraic  topology and persistent homology is  \cite{edelsbrunner2010}.

%%%% SECTION 2 %%%%%%%%%%
%%%%%%%%%%%%%%%%%%%%5

\subsection{Vietoris-Rips complex and topological descriptors }\label{sec:methods}
We will further explain Figure \ref{fig:doubleAnn} and then introduce three topological descriptors, barcode, {persistence diagram}~\cite{Edels2002} and {persistence landscape}~\cite{bubenik2012}.
A collection of point cloud data in a metric space is converted to  a combinatorial graph whose edges are determined by closeness between the points. 
While a  graph captures  connectivity and clustering of data, it  ignores higher dimensional features.
This  idea of graph can be extended to a {simplicial complex}, which is a collection of simplices.

Suppose there is a finite set of points $\{v_i\}_1^n$  in $\reald$ and (dis-)similarity measure is denoted as $d(v_i, v_j)$.
A $k-$simplex is  a set of all points $x\in \reald$ such that $x=\sum_{i=1}^{k}, a_iv_i$, where $\sum a_i=1, a_i \geq 0.$
It is easy to picture low dimensional simplices; a 0-simplex a vertex, a 1-simplex an edge joining two vertices,  a 2-simplex a triangle, 
 a 3-simplex a tetrahedron.
\emph{Vietoris-Rips} complex $\vr_{\eps}$ is a set of simplices whose vertices have pairwise  distance within $d(v_i, v_j) \leq \eps.$
Algebraic topology adds group structure onto the complex,  $H_k(\vr_{\eps}),$ called $k-$th {homology group}. 
For coefficients in a field, $H_k(\vr_{\eps})$ is a vector space, whose basis consist of linearly independent $k-$dimensional cycles that are not  boundaries. 
The $k-$th Betti number $\beta_k$, is the rank of $H_k(\vr_{\eps})$ and  counts the number of $k-$dimensional holes of a simplcial complex. 

At each fixed $\eps$, homology group $H_k(\vr_{\eps})$  can be calculated, but  computational topologists think of {persistence}.  The simplical complexes grow as $\eps$ increases, that is,  $\mathcal{V}_{\eps} \subseteq \mathcal{V}_{\eps*},$ for $\eps \leq \eps^*,$ this inclusion induces a linear map,  $H_k(\vr_{\eps}) \rightarrow H_k(\vr_{\eps^*}).$ This allows us to examine the filtration of homology.   
The evolution of the simplicial complexes over  increasing values of $\eps$ can be completely  tracked using barcodes or  persistence diagrams.  
Barcode is the  multiset of intervals $(\eps', \eps'')$, where $\eps'$  and $\eps''$ indicate birth and death time of a  topological feature. 
Alternatively,  the birth and death times can be represented by a point $(\eps',\eps'')$ in $\mathbb{R}^2$. The collection of these points in $\mathbb{R}^2$ is a persistence diagram (see Figure \ref{fig:pd_10}). Vietoris-Rips complex and its evolution are demonstrated with random points selected from a double annulus, see Figure \ref{fig:doubleAnn} above. 

\begin{figure}[htp]
  \begin{center}
     \includegraphics[scale=0.58]{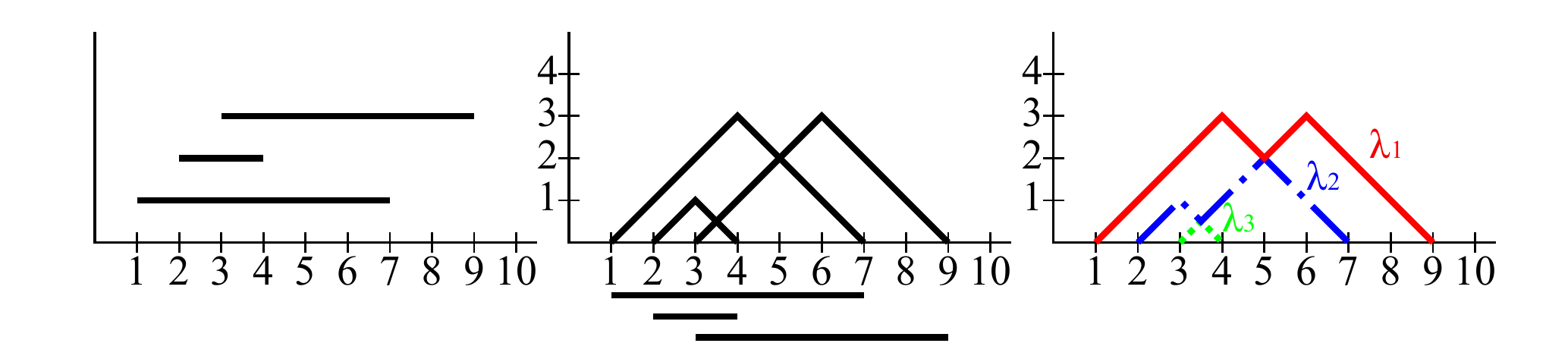}
  \end{center}
  \caption{(Left) Three intervals in a barcode. (Middle) Isosceles triangles of intervals. (Right) Persistence landscape.}
  \label{fig:pl_constr}
\end{figure}

In a simple summary, the point cloud data are transformed to barcodes or  persistence diagrams in each  dimension.
Is it possible to calculate means and variances of barcodes or persistence diagrams?
It is well known that  the Fr\'{e}chet mean of barcode (persistence diagram) is not unique.  
Many researchers  have advanced in developing  theories  in this research area~\cite{fasy2013, mileyko2011}.  
Bubenik~\cite{bubenik2012} introduced  a third topological descriptor, persistence landscape.  
Given an interval $(b, d),$ with $ b\leq d,$ define a function leading to an isosceles triangle,  $  f_{(b,d)}: \real \rightarrow \real, 
f_{(b, d)}=\min (t-b, d-t)_{+},$  where $u_+=\max(u,0).$ 
The persistence landscape  corresponds to a multiset of intervals $\{(b_i, d_i) : b_i \leq d_i\}$ and to a set of functions, 
$\{ \lambda(k, t): \n\times  \real \rightarrow \real \}$ where : $\lambda (k, t)$ is the  $k-$th largest value of $\{f_{(b_i, d_i )}\}.$
See an illustration in Figure \ref{fig:pl_constr}.

% figure PLs
%%% FIGURE 2%%%%%%%%%%%

%%%% SECTION 4 %%%%%%%%%%
%%%%%%%%%%%%%%%%%%%%
\subsection{ Statistical Inference with Persistence Landscape} \label{sec:PL}
On $\n \times \real$, the product of the counting measure on $\n$ and the Lebesgue measure on $\real$ are used.
The persistence landscape,  $\lambda(k,t): \n\times \real \rightarrow \real$,  is bounded and nonzero on a bounded domain.
Hence persistence landscape belongs to $\pl^p(\n\times \real)$, with a  metric induced by $p$-integrable functions and hence  is a separable Banach space~\cite{bubenik2012}.
Bubenik also showed that when $p \geq 2$, with finite first and second  moments, persistence landscape satisfies a Strong Law of Large Numbers (SLLN) and a Central Limit Theorem (CLT).
Suppose $\Lambda_1, \ldots, \Lambda_n$ are the random variables corresponding  to persistence landscapes. The vector space structure of $\pl^p(\n\times \real)$ induces  the mean 
landscape as the pointwise mean, $\bar{\lambda}(k,t)=\frac{1}{n} \sum_{i=1}^{n} \lambda_i(k, t).$

On $p$-integrable space,  $\pl^p(\n\times \real),$ where $ p \geq 2$  and under the assumption of  finite first and second moments, 
for any continuous linear functional $f$, the random variable $f(\lambda(k,t))$ also satisfies  SLLN and CLT~\cite{Ledoux2002}.
There are many choices of $f$, but the integration of $\lambda$ might be a natural choice,
$f(\lambda(k,t))=\sum\limits_{k} \int_{\real} \lambda(k, t)\,dt.$ It has a good interpretation; the  values  of $f$ 
are an enclosed total  area of all curves  $\lambda(k,t).$ Choice of $f$ might depend on  data, see other choices in \cite{bubenik2012}.   
With the integration as a functional choice, 
we are ready to set hypotheses.
Let $Y_1=f(\lambda_1(k,t))$ and $Y_2=f(\lambda_2(k,t))$ be random variables for groups 1 and 2.
We let $\mu_1$ and $\mu_2$ be corresponding population means. The hypothesis of interest is
\begin{equation}
H_0: \mu_1-\mu_2=0  \;\; \mbox{vs.} \;\; H_a: \mu_1- \mu_2 \neq 0.
\label{eq:hypo}
\end{equation}

We also consider the similarity measure between persistence landscapes  as well as between persistence diagrams.
The measure between persistence landscape is defined as the $p$-norm of difference.
Suppose there are two samples that have landscapes denoted as $\lambda_1(k, t)$ and $\lambda_2(k, t)$. Then the $L_p$-norm is defined in (\ref{eq:PLdiff}). This compares pairwise the area under the contours between $\lambda_1(k,t)$ and $\lambda_2(k,t)$

%FORMULA FOR DIFFERENCE BETWEEN PERSISTENCE LANDSCAPES
\begin{equation}
||\lambda_1-\lambda_{2}||_p=\left(\sum\limits_{k}\int_\mathbb{R}|\lambda_1(k,t)-\lambda_2(k,t)|^p\right)^{1/p}.
\label{eq:PLdiff}
\end{equation}
For our data, we will calculate the persistence landscape  distance using both $L_1$ and $L_2$ norm in  (\ref{eq:PLdiff}).
Wasserstein distance (see  (\cite{edelsbrunner2010}, for example) is a popular  measure of dissimilarity  between persistence diagrams.
All the results based on persistance landscape ($p=1, 2$)  and Wasserstein distance are similar and so all our statistics and presentations are  based on $L_2$ norm in  the following sections. 

%%%% SECTION 5
%%%%%%%%%%%%%%%%%

\section{Topological data analysis}\label{sec:avgPL}

\begin{figure}[htp!]
 \begin{center}
\begin{tabular}{c}
    \subfigure[Degree 0-patient  10]{\label{fig:edge-a}\includegraphics[scale=0.3]{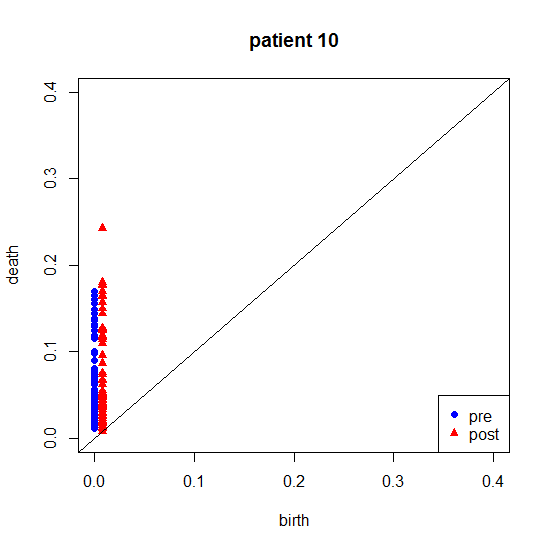}} 
    \subfigure[Degree 1-patient 10]{\label{fig:edge-b}\includegraphics[scale=0.3]{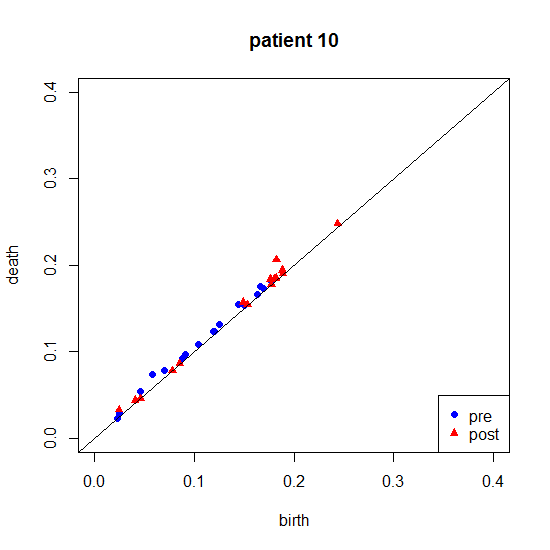}} \\
\end{tabular}
  \end{center}
  \caption{Persistence diagrams for patient 10 before (in blue)  and after (in red)  FMT in (a) degrees zero  and (b) one. Note that the triangles in (a)  have birth at time zero but are moved slightly for visual purposes.}
  \label{fig:pd_10}
\end{figure}

The computation of the Vietoris-Rips complex, which is computationally intensive, is  carried out using the \texttt{phom} package in R \cite{tausz2011} on the \textit{Westgrid} computer network. 
All 45 samples had intervals in barcode in degrees zero and one. However in degree two, only 12 of the pre-FMT samples and 15 of the post-FMT samples have intervals in barcode. Persistence diagram  makes visual pairwise comparisons easier than barcode. 
Figure \ref{fig:pd_10} shows the corresponding persistence diagrams of patient 10 before and after the FMT  treatment. 
%%% FIGURE 3%%%%%%%%%%%
%\newpage
At first glance there does not appear to be anything interesting but there are some general trends. We observe that the degree 0 persistence  diagram for the pre-FMT samples have components that die a lot sooner than those in the post-FMT samples. Similarly, the birth and death times of these loops are shorter for the pre-FMT samples than for the post-FMT samples.
These observations  may indicate that there could be a true difference in the topological structure in degrees zero and one between the two groups. On the other hand,  many of the intervals in degree one  are very short, thus what we observe may be noise rather than signal. 

Unlike barcode or persistence diagrams, we can calculate means and variances of persistence landscapes.
Figure \ref{fig:avgPL} below, shows the average persistence landscapes of pre-FMT and post-FMT samples in degree 0 and 1. 
The trend in average persistence landscapes is the same as in persistence diagrams or barcodes, that is, the post-FMT samples have slightly longer intervals than the pre-FMT in degree 0 and 1.
%%% FIGURE  4 %%%%%%%%%%%
\begin{figure}[htp!]
 \begin{center}
\begin{tabular}{c}
    \subfigure[pre-FMT-degree 0]{\label{fig:edge-a}\includegraphics[scale=0.2]{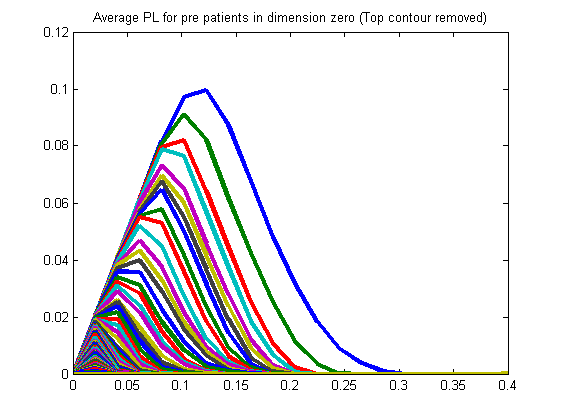}} 
    \subfigure[post-FMT-degree 0]{\label{fig:edge-b}\includegraphics[scale=0.2]{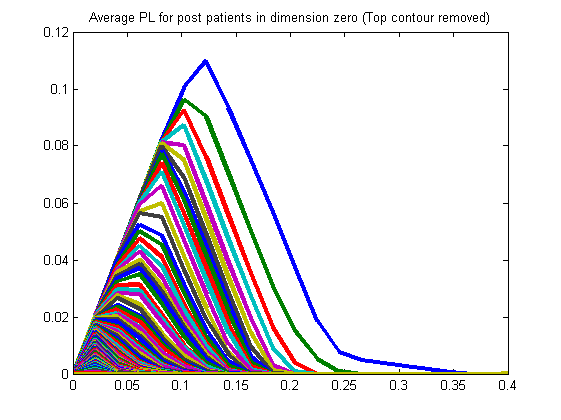}} 
    \subfigure[pre-FMT-degree 1]{\label{fig:edge-c}\includegraphics[scale=0.2]{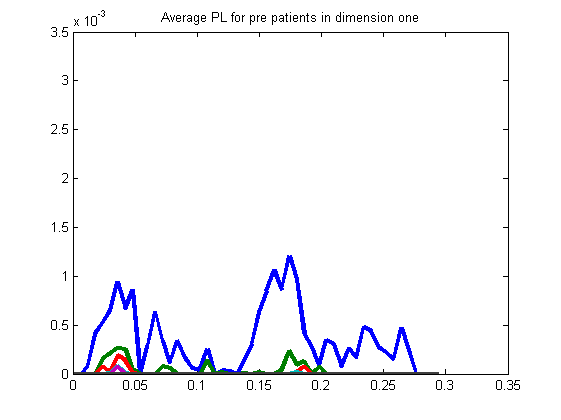}} 
    \subfigure[post-FMT-degree  1]{\label{fig:edge-d}\includegraphics[scale=0.2]{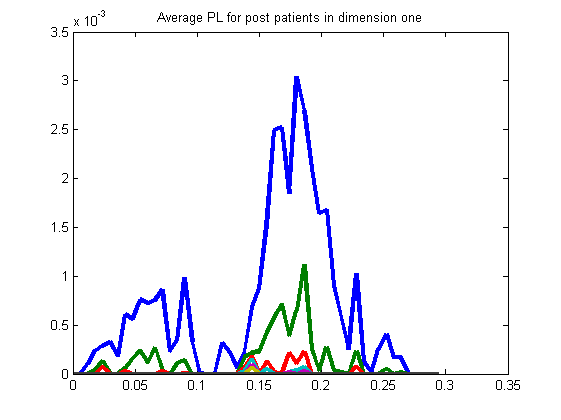}} \\
\end{tabular}
  \end{center}
  \caption{Average persistence landscapes for pre-FMT and post-FMT patients in degrees zero and one. In degree zero, the post-FMT group has a denser grouping of contours, which means that there are more clusters than in the post-FMT samples. In degree one, the pre-FMT group has three persistent loops on average and the post-FMT sample has two. However, the post-FMT sample loops are more persistent than the pre-FMT samples. We note that the vertical scales in the plots in degree 0 and 1 are different.}
  \label{fig:avgPL}
\end{figure}

Quadratic discriminant analysis~(QDA)  was performed on  the $\beta_0$- and $\beta_1$-Isomap embedded coordinates.
Three groups are well separated on the plane of both degree 0 and 1. 
It is interesting to see that the donors are grouped on one side and pre-FMT patients  on the other side, and  post-FMT patients between the two. 
This is in complete agreement with what clinicians believe is happening and it is frequently reported that patients gut microbiome take on the characteristics of the donor microbiome
following an FMT, see \cite{Shahinas, weingarden2014, khanna2016}. 

\begin{figure}[htp!]
 \begin{center}
\begin{tabular}{c}
\subfigure[Degree 0]{\label{figZhai0}\includegraphics[scale=0.40]{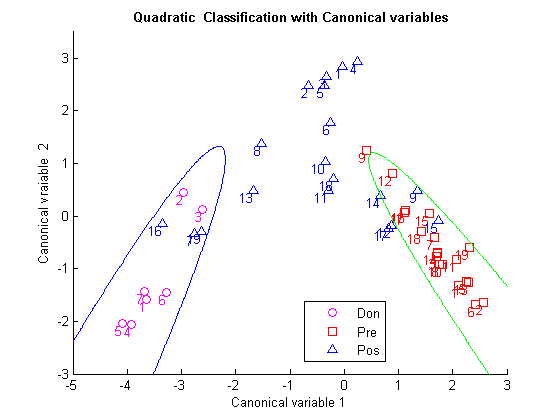}} 
\subfigure[Degree 1]{\label{figZhai1}\includegraphics[scale=0.40]{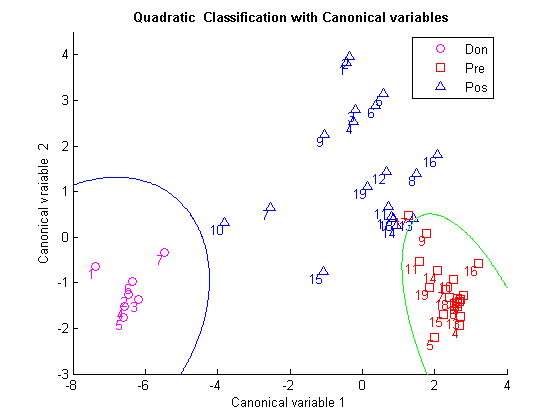}} \\
\end{tabular}
\end{center}
\caption{ Classification in two-dimensional space. Three groups are well separated on the $\real^2$ for both  degree 0 and 1.
The canonical functions  for both $\beta_0$- and $\beta_1$- Isomap discriminate pre-FMT  patients and donors  on each quadrant while post-FMT patients 
located between the two. 
(Left) The patients 7, 16 and 19 after FMT treatment classified as donors, possibly indicating  their donors might be 2 and 3. 
Patients 9 and 15 do not appear `improved' after FMT as they are classified as the group of patients before the treatment.
The patients 7, 16 and 19 are not only similar to donors but also further away their pre-FMT position.
This may indicate they have improved the most after the treatment.
(Right) In terms of  loops in DNA sequences,  a cluster of  post-FMT patients are similar to the group of post-FMT treated patients. 
Donor 7 could be the donor for patient 10.}
\label{fig:PLscatter}
\end{figure}

It would have been very interesting to compare the matched donor and the patient after FMT, but this information was not recorded during this study, \cite{lee2014}.
The classification in $\mathbb{R}^2$ space  in degree 0  in Figure \ref{fig:PLscatter} shows the patients 7, 16 and 19 post-FMT, become much like donors, particularly  donors 2 and 3. Patients  9 and 15 post-FMT  did not seem improved as they are similar to the cases pre-FMT.
The classification in $\mathbb{R}^2$ space  in degree 1  in \ref{fig:PLscatter}  show a few post-FMT patients  (7, 10, 15) are close to the donors after FMT while most post-FMT patients remain similar to  pre-FMT.  Following \cite{lee2014}, a clinical trial comparing the efficacy of frozen {\it versus} fresh FMT has been completed, see \cite{lee2016}.  Here stool samples were
collected at pre-FMT, followed by
day-10, week-5, week-13 following a patients last FMT along with the exact donor stool sample pairing.  Sequencing of this data is currently underway.

%%% FIGURE  5 %%%%%%%%%%%

We recall that our `raw' data was the dissimilarity matrix between 147  uniques DNA sequences {\it per} subject, so a total of forty five $147\times 147$
matrices. For each matrix, we construct a Vietoris-Rips complex, then calculate  persistence landscapes which enables us to perform statistical inference. 
Hypothesis  test (\ref{eq:hypo}) were carried out to comparing pre-FMT and post-FMT samples. 
The $p-$values of the paired $t-$tests are 0.0064, 0.0083 and 0.2591 in degree 0, 1, and 2, respectively.
Since all the analysis above is based on 147 sequences randomly chosen from  those patients whose number of  of DNA sequences are bigger than 147,   we repeated analysis 10 times with  independent 147 DNA  samples. The test statistics  and $p-$values for 10 runs are similar showing consistency of result regardless of which  147 DNA sequences were applied. 

Hypothesis tests  show significant difference  of topological features in degree 0 and 1 between patients before and after FMT treatment.  
For degree 0, this implies that  the number of  clusters  and their persistence on DNA sequences in pre-FMT samples  are different from  those in post-FMT samples. For degree 1, this implies that  the  number  of loops (cycles) and their persistence in DNA sequences in pre-FMT samples  are different from  those post-FMT samples.
We  present  DNA sequences  as points  cloud in $\real^3$ and observe  patterns of clustering and loops in DNA sequences in  the following section.

%%%% SECTION 6%%%%%%%%%%
%%%%%%%%%%%%%%%%%%%%
\subsection{Clusters and loops in  DNA sequences}\label{sec:tdaDNA}

Applying dimensional reduction methods; Isomap and multidimensional scaling (MDS), to dissimilarity measure between DNA sequences, we project the DNA sequences to $\real^3$ and obtain embedded coordinates.
Scree plots appear to indicate embedding dimension of the DNA sequences is 3.
The  residual variance  in MDS was much higher than for Isomap, hence figures based on Isomap  are presented below.

%%% Figure 6%%%%%%%%%%%5
\begin{figure}[!htbp]
  \begin{center}
\subfigure[Donor 3]{\label{fig:edge-a}\includegraphics[scale=0.30]{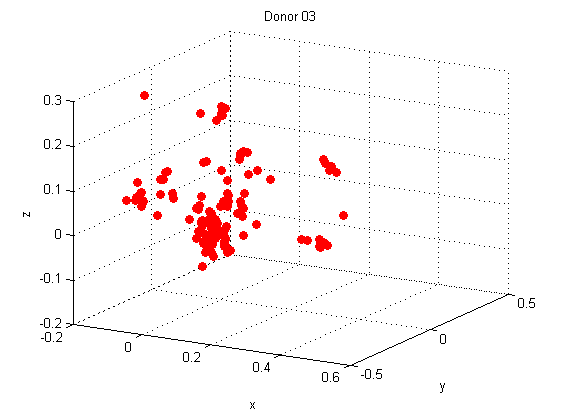}} 
\subfigure[Post 7]{\label{fig:edge-b}\includegraphics[scale=0.30]{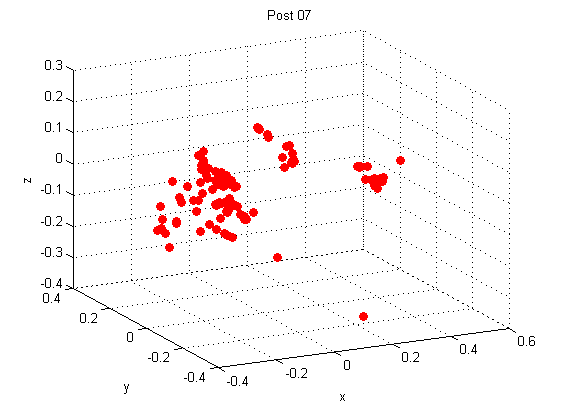}} 
\subfigure[Pre 19]{\label{fig:edge-c} \includegraphics[scale=0.30]{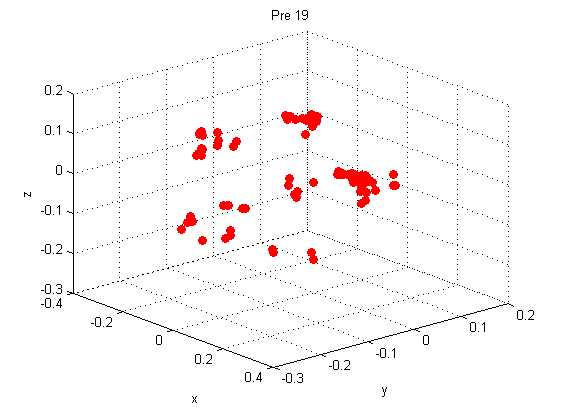}}
\subfigure[Post 19]{\label{fig:edge-d}\includegraphics[scale=0.30]{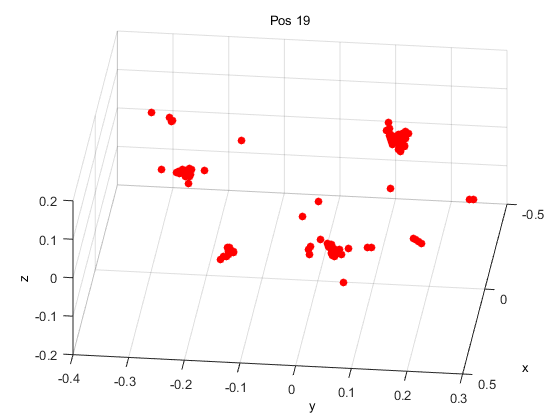}}
  \end{center}
  \caption{
Plots (a)  and (b):  147 DNA sequences  donor 3  and post-FMT patient 7 on Isomap  embedded coordinates. Both samples have roughly
 the same number of clusters. In general, donors and patients that were close on the classification in $\mathbb{R}^2$ in Figure  \ref{fig:PLscatter}
 had similar number of clusters.
Plots (c) and (d)  147 DNA sequences  of pre-FMT and and post-FMT patient 19  on Isomap  embedded coordinates. 
There are several clusters in both  pre-FMT/post-FMT patient 19.  The number of clusters in DNA sequences of post-FMT patient 19 is higher, but there are many clusters  formed by a single DNA.
}
\label{fig:scatter_betti0}
\end{figure}

Figure~\ref{fig:scatter_betti0} shows the Isomap embedded coordinates for donor 3, post-FMT patient 7 and pre-FMT/post-FMT patient 19.
 From the Figure~\ref{fig:scatter_betti0}, it can be seen that donor 3 and post-FMT patient 7 have a similar spread among the sequences and there are 2-3 distinct clusters. 
We have also  noticed that donor 2 and post-FMT patient 18 also have similar structures. 
The donors generally have a wider spread, which would indicate more diverse DNA sequences and hence  a healthier gut microbiome.
For the number of of clusters, there are higher number of clusters in post-FMT samples,  however  some clusters contain only one DNA sequence  and  are more spread out.
For example, there are about 5 groups in  pre-FMT patient 19;   many clusters with fewer DNA sequences in each cluster in post-FMT patient 19.  
Ignoring the singleton clusters, we observe two  large clusters in post-FMT patient 19.
 This information was  shown  in Figure \ref{fig:scatter_betti0} as well as on  barcode and persistence diagram.  The two clusters in post-FMT patient 19 can be seen in the barcode diagram as the two points that have the highest €`death'€™ time. The spread out clusters in pre-FMT 19 are shown as the bars that have the earlier birth and earlier death times
(figures are not shown here). This trend is visible in other samples and this might explain the small $p-$value for testing the difference in area under the persistence landscapes which was calculated in Section \ref{sec:avgPL}.

%%%% SECTION 7 %%%%%%%%%%
%%%%%%%%%%%%%%%%%%%%

\section{Conclusions} \label{sec:conclusion}
We illustrated how topological data analysis can be applied to similarity measures data of DNA sequences.
DNA sequence data was analyzed using three summary statistics of persistent homology; namely, barcodes, persistence diagrams and persistence landscapes. The main objective was to see if there are any differences in the topological feature of DNA sequences in the gut microbiome of CDI patients before and after FMT treatment. 

From visual inspection of barcodes and persistence diagrams it was seen that the components in dimensions zero and one died sooner in the pre-FMT samples than in the post-FMT samples. Persistence landscapes were able to present this difference more formally, showing that there is a difference in the average area under the persistence landscapes for pre-FMT and post samples. Alternative interpretation of this was that there was a difference in the number and size of clusters in dimension zero, and the number and size  of loops in dimension one. The post-FMT  samples had more clusters than the post  samples, whereas there was no visually obvious difference in the size of the loops, but the loops were bigger for the post samples.

We performed discriminant analysis on $\beta_0, \beta_1$ -Isomap embedded coordinates.
The classification on two dimensional space for both degree 0 and 1  show good separation among three groups.
For degree 0, the first two  discriminant function  separates  pre-FMT  patients and donors, and post-FMT patients sit  between the two.
For degree 1, the first two  discriminant function  separates  pre-FMT and post-FMT patients and donors are between  the two.

The major drawbacks of this analysis is that information about individual sequences is lost. This project looked at the topological structure created by the sequences but no details were provided about the individual sequences. As the micro-biological technology and methods improve it may be interesting to incorporate this information.  Also of interest would be meaningful identification of bacterial species. Studies of the 16S rRNA gene only measure presence of the species, but do not say anything about their functionality.  For the latter one has to turn to the metabolome which is currently
under investigation with the patients in \cite{lee2016}.

\begin{acknowledgement}
We are grateful to all of the donors, families and patients 
who took part in this study.  We also appreciate the clinical and research staffs at St Joseph's Healthcare Hamilton
where the clinical work had been performed.  The corresponding
author would also like to thank the participants of the SAMSI Working Group ``Nonlinear 
Low-dimensional Structures in High-dimensions for Biological Data" which was part of the 2013-14 SAMSI LDHD Program.  Much of the
discussions were centred on the work presented.

We  also thank Violeta Kovacev-Nikolic for her help with matlab code and Figure 1;  Professor Patrick Schloss for his help using {\tt mothur}; and Yi Zhou for his help with Figures 2 and 3.
Computations in this research were largely enabled by resources provided by WestGrid and Compute Canada.

We would like to acknowledge funding support provided by:  CANSII CRT;  CIHR 413548-2012; McIntyre Memorial Fund; Michael Smith Foundation; NSERC DG 293180, 46204; NSF DMS-1127914; and, PSI Foundation Health Research Grant 2013, 2017.

The study and permission protocol was approved by the Hamilton Integrated
Research Ethics Board \#12-3683, the University of Guelph
Research Ethics Board 12AU013 and the University of Alberta,  Health Ethics approval Pro00047221.
\end{acknowledgement}

%\bibliographystyle{plos2009}
%\bibliographystyle{vancouver}

%\bibliography{\bibliographystyle{plos2009}

\bibliographystyle{abbrvnat}
\bibliography{bibliography.bib,PhyLASSO.bib,cdiff.bib,thesis.bib}

\end{document}